\documentclass[dvips]{acta}
\usepackage{supertabular,lscape,epsfig}
\usepackage{amssymb}
\usepackage{amsmath}

\SetPages{1}{1}

\SetVol{65}{2015}

\def\figfld{1} 
\def\figfldcore{2} 
\def\figstddev{3} 
\def\figismcomp{4} 
\def\figmaxamp{5} 
\def\figlcwvir{6} 
\def\figlcrrab{7} 
\def\figlcrrc{8} 
\def\figlcsxphe{9} 
\def\figlcother{10} 
\def\figvarcharts{11} 
\def\figcmd{12} 
\def\figcmdhst{13} 
\def\figmultivix{14} 
\def\figmultivxviii{15} 
\def\figsxpheppr{16} 
\def\figsxpheplr{17} 
\def\tabvarpar{1} 
\def\tabsxphedm{2} 

\def\[#1]{{\bf #1}}

\begin{document}

\begin{Titlepage}
\Title{Wide-field Variability Survey of the Globular Cluster M\,79 and a New Period -- Luminosity Relation for SX Phe Stars}
\Author{G. K~o~p~a~c~k~i}%
       {Instytut Astronomiczny, Uniwersytet Wroc\l{}awski,
        Kopernika 11, 51-622 Wroc\l{}aw, Poland\\
        e-mail: kopacki@astro.uni.wroc.pl}

\Received{Month Day, Year}
\end{Titlepage}

\Abstract{%
 We present the results of a search for variable stars in a 26$\times$39 arcmin$^{\rm 2}$ field around globular
 cluster M\,79 (NGC\,1904). The search was made by means of an extended version of image subtraction, 
 which allows to analyze in a uniform manner CCD frames 
 obtained with different telescopes and cameras of different sizes and resolutions.
 The search resulted in finding 20 new variable stars, among which 13 are cluster members. The members include 
 one new RR Lyr star of subtype c, three SX Phe stars, and
 nine variable red giants. We also show that V7 is a W Vir star with a period of
 13.985 d.
 Revised mean periods of RRab and RRc stars, 
 $\langle P\rangle_{\rm ab}={}$0.71~d and $\langle P\rangle_{\rm c}={}$0.34 d, respectively, and
 relative percentage of RRc stars, $N_{\rm c}/(N_{\rm ab}+N_{\rm c})={}$45 \%
 confirm that M\,79 belongs to the Oosterhoff II group of globular
 clusters.
 The mean $V$ magnitude of the horizontal branch of M\,79 based on ten RR Lyr stars
 has been estimated to be $V_{\rm HB}=\langle V\rangle_{\rm RR} ={}$16.11${}\pm{}$0.03 mag.
 In one RRc star, V9, light changes with three close frequencies were detected, 
 indicating excitation of  nonradial modes. An SX Phe star, V18, is a double-mode
 pulsator with two radial modes excited, fundamental and first overtone. 
 Moreover, we have discovered two SX Phe or $\delta$ Sct stars 
 and one W UMa type system, all likely field objects.
 We also studied the 
 period -- luminosity relation for SX Phe stars. Using 62 
 fundamental and fundamentalized periods of radial double-mode and high-amplitude SX Phe stars known in 
 Galactic globular clusters, we have derived the slope and zero point of this
 relation to be, $-$3.35${}\pm{}$0.27 and 2.68${}\pm{}$0.03 mag (at $\log(P/{\rm d})=-$1.24), respectively.
 }%
 {stars: Population II -- 
  stars: RR Lyr variables -- 
  stars: SX Phe variables -- 
  globular clusters: individual: M\,79}

\section{Introduction}

 Globular clusters represent the oldest, and consequently metal poor, ingredients of the Galaxy.
 Their study is thus very important in the context of formation and early
 evolution of the Milky Way.
 They were usually considered a good example of 
 simple stellar population, in the sense that all cluster
 members share the same age and initial chemical composition,
 differing virtually only in mass. Recent progress in studies
 of globular clusters show, however, that they are more complex objects.
 In particular, they can consist of multiple stellar populations; 
 see Gratton {\em et al.\/} (2012) for a review.
 
 Our ongoing observational project is aimed at detecting and analyzing
 RR Lyr and SX Phe type pulsating stars in globular clusters (Kopacki 2000, 2001, 2005, 2007, 2013, 2014). 
 The overall properties of RR Lyr
 stars as a group are related to the intrinsic parameters of the host
 cluster, such as metallicity and age. This manifests in a division of
 Galactic globular clusters into two Oosterhoff types distinguished by the
 average period of their RRab stars (see Clement and Rowe 2000 and references therein). 
 SX Phe stars, which are blue stragglers in globular clusters, 
 are interesting because those pulsating in radial modes
 follow period -- luminosity relations (Nemec {\em et al.\/} 1994, McNamara 1995,
 Cohen and Sarajedini 2012). 
 These stars can be therefore used as independent
 distance indicators. 
 
 In this paper we present the results of a variability survey in M\,79.
 Preliminary results of this analysis were already 
 published by Kopacki and Pigulski (2012a). 
 Moreover we discuss period -- luminosity relation for SX Phe
 stars based on radial double-mode stars of this type known in Galactic globular
 clusters.

\section{The Cluster}

 M\,79 (NGC\,1904, C0522-245 in the IAU nomenclature) is a southern globular cluster  with an
 intermediate metallicity [Fe/H]${}={}-$1.59 according to the updated Harris' (1996) catalogue.%
 \footnote{http://physwww.physics.mcmaster.ca/~harris/mwgc.dat}
 The structural parameters of M\,79, taken from the same catalogue, are: the core radius 
 $r_{\rm c}={}$9.6 arcsec and the half-light radius $r_{\rm h}={}$39 arcsec.
 The cluster's tidal radius determined by Lanzoni {\em et al.} (2007) equals to $r_{\rm t}={}$8.3 arcmin.
 Color -- magnitude diagram of the cluster shows prominent extended blue horizontal branch
 (hereafter BHB), 
 very similar to what is observed for M\,13 (Kravtsov {\em et al.\/}\ 1997, Dalessandro {\em et al.\/}\ 2013).
 In the deep photometric study of Lanzoni {\em et al.\/}\ (2007), based on the HST observations,
 the cluster was shown to contain a relatively large and centrally concentrated
 population of blue stragglers. M\,79, along with NGC\,1851, NGC\,2298, and NGC\,2808, was
 suggested by Martin {\em et al.\/} (2004) to be associated with the Canis Major (CMa) dwarf galaxy.
 This idea was, however, questioned by other studies ({\em e.g.\/} Mateu {\em et al.\/} 2009) due to the
 lack of a BHB stars and no excess of RR Lyr stars in the CMa overdensity.

 The most recent version of the Catalogue of Variable Stars in Globular Clusters%
 \footnote{http://www.astro.utoronto.ca/~cclement/cat/listngc.html} 
 (CVSGC, Clement {\em et al.\/}\ 2001) lists
 14 variable objects in the field of M\,79 including ten RR Lyr stars,
 V3 -- V5 (Bailey 1902), V6 (Rosino 1952), V9 -- V13 (Amigo {\em et al.\/} 2011), and
 V14 (Kains {\em et al.\/} 2012). The remaing four stars exhibit mosty long-term
 variability. Two of them, V7 and V8 have no firm classification assigned, although
 V7 is suspected to be Pop.\ II Cepheid with an approximate period of 0.7 d.
 We note here, that variables V9 through V14 were independently discovered by
 Kopacki and Pigulski (2012a).

\begin{figure}[h!tb]
\hbox to\hsize{\hss\includegraphics{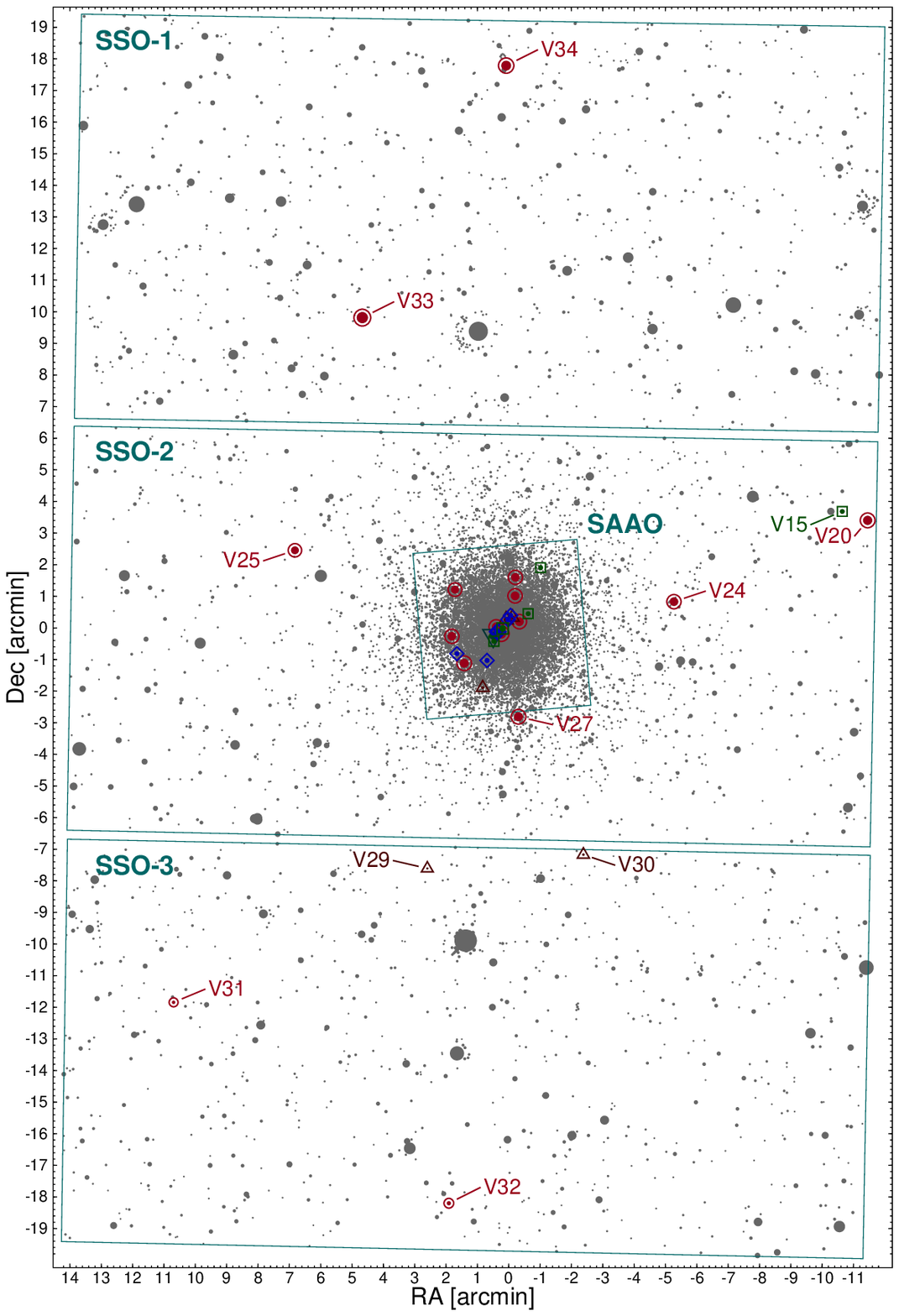}\hss}
\FigCap{%
 Schematic view of the observed fields of M\,79 (SSO-1,2,3 and SAAO) covering
 altogether 25.5${}\times{}$38.8 arcmin$^2$. 
 Variable stars are indicated with different symbols,
 a W Vir star, with a bottom-tipped triangle, RRab variables, with diamonds,
 RRc stars, with squares, SX Phe stars, with top-tipped triangles, 
 and stars of other types (eclipsing systems, variable red giants, unknown),
 using open circles. Only variable stars located outside the SAAO field are labeled.
 The $(0,0)$ coordinates correspond to 
 $\alpha_{2000}={}$5$^{\rm h}$24$^{\rm m}$10$^{\rm s}$, 
 $\delta_{2000}={}-$24$^\circ$31$^\prime$27$^{\prime\prime}$.}
\end{figure}

\section{Observations}

 The CCD observations presented here were obtained at the 
 South African Astronomical Observatory (SAAO),
 Sutherland, Republic of South Africa, and at the Siding Spring
 Observatory (SSO), Coonabarabran, Australia. At SAAO we
 used the 40-inch telescope equipped with a STE4 detector. It was 
 a SITe back-illuminated CCD camera with a size of 1024$\times$1024 px (`px' stands for `pixel'), 
 gain of 2.8 e$^-$/ADU and a read-out noise of 6.5 e$^-$. At SSO we made use
 of the 40-inch ANU telescope equipped with the Wide Field Imager (WFI). Only three chips of
 the WFI were utilized (their alignment is shown in Fig.\ \figfld). 
 Each of them had a size of 4096$\times$2048 px,
 gain of 4.7 e$^-$/ADU and an average read-out noise of 7.2 e$^-$.

 
 The observations were carried out in 2008, from February 2 to March 17 at SSO, and 
 from April 9 to 15 at SAAO. Since our main goal was a search for
 variable stars in M\,79, we observed mostly through $V$ filter of the Johnson-Cousins $UBV(RI)_{\rm C}$ system,
 but to secure color infromation we observed also in the $I_{\rm C}$ passband (only at SSO). 
 In total, we obtained 769 $V$-filter and 255 $I_{\rm C}$-filter CCD frames of the cluster. 

 Each CCD chip of the SSO WFI camera covered an area of 25.7$\times$12.8 arcmin$^{\rm 2}$, 
 thus one integration with three chips covered about 26$\times$39 arcmin$^{\rm 2}$.
 Hereafter, we designate the three subfields of the WFI as SSO-1, SSO-2, and SSO-3, with the first
 one being the northernmost.
 The frames obtained at the SAAO covered about 5.2$\times$5.2 arcmin$^{\rm 2}$.
 The outlines of the observed fields are shown in Fig.\ \figfld. As can be seen, the cluster
 core was placed in the center of both SSO (in particular SSO-2) and SAAO fields.

 The exposure times amounted to
 300 s in the $V$ filter, and to 200 s in the $I_{\rm C}$ filter. 
 On most nights the weather was very good but some nights were interrupted by 
 a strong wind and, especially in the morning, by a thick mist. 
 The seeing was in the range from 1.1 to 5.0 arcsec and from 1.3 to 2.7 arcsec, 
 for SSO and SAAO observations, respectively.

\begin{figure}[h!tb]
\hbox to\hsize{\hss\includegraphics{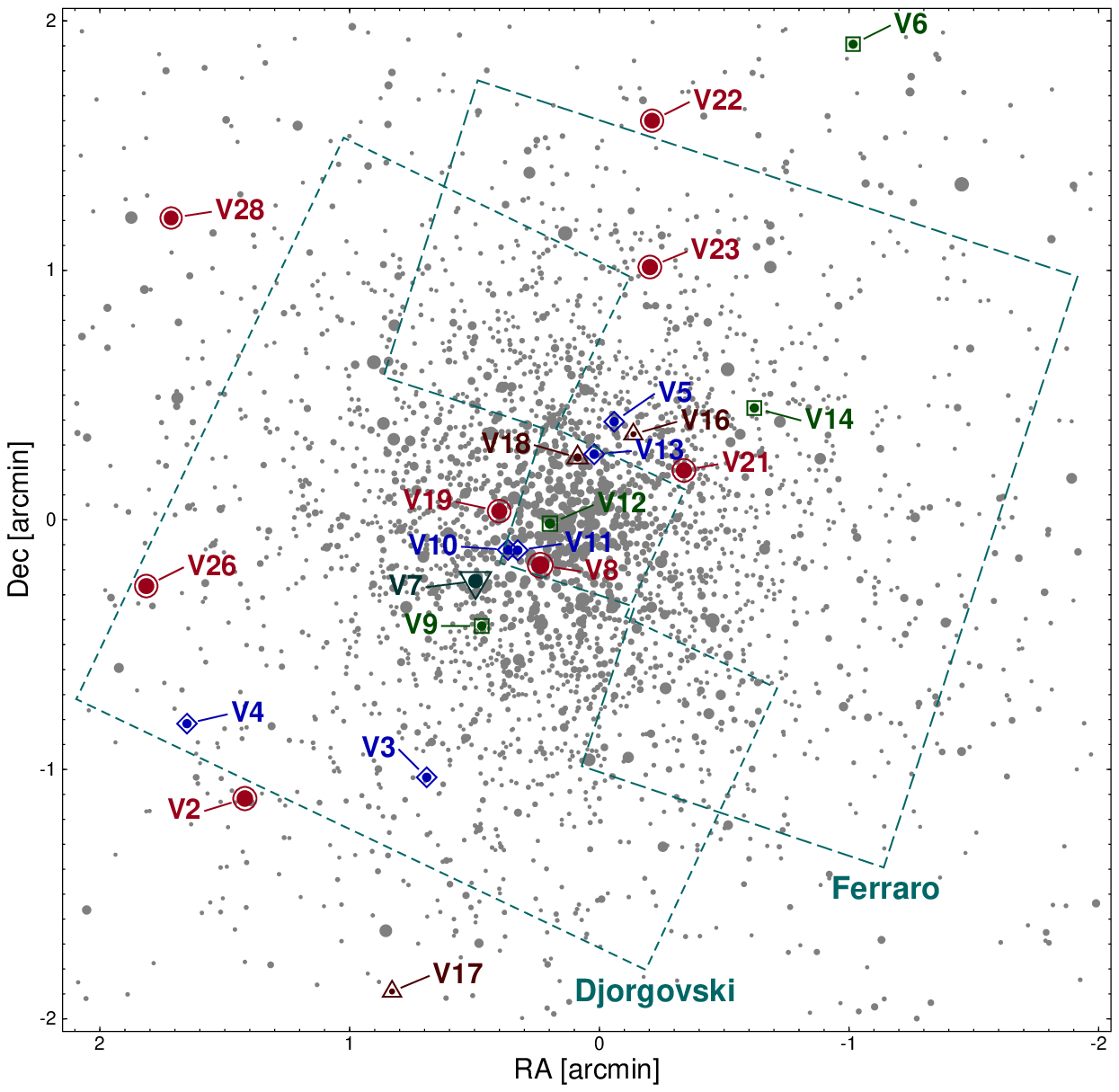}\hss}
\FigCap{%
 The close-up of the previous figure showing central region of M\,79 and covering
 4${}\times{}$4 arcmin$^2$. The outlines of the two HST WFPC2 pointings 
 (PIs: Djorgovski (GO\,6095) and Ferraro (GO\,6607)) are shown
 with dashed lines.
 Variable stars are indicated with different symbols, using the same scheme as in 
 Fig.\ \figfld, and labeled with their numbers given
 in Tab.\ \tabvarpar.}
\end{figure}

\section{Photometric Reductions}
  
 The pre-processing of the CCD frames was performed in the usual way and 
 consisted in a 2${}\times{}$2 binning (only for the SSO data), subtraction of the 
 overscan and mean bias frame and applying flat-field correction using mean flat-field
 frames constructed for each night separately. The dark current was found to be negligible
 for both instruments. Instrumental magnitudes for all stars in the observed fields were 
 derived with the DAOPHOT profile-fitting package of Stetson (1987). All images were 
 reduced using the method described by Jerzykiewicz {\em et al.\/}\ (1996). We identified 
 about 4200, 16500, and 2700 stars in the observed three SSO subfields, respectively, 
 but could not resolve well the dense cluster core. A schematic views of the monitored fields are presented
 in Figs.\ \figfld\ and \figfldcore.

The differential photometry was derived on the frame-to-frame basis
rather than on the usual star-to-star basis, using the procedure
described by Kopacki {\em et al.} (2008). In short, before
extracting light curves for every observed star, 
instrumental photometry derived from each frame was shifted to the same
magnitude scale (defined by means of a reference frame) 
by an average offset in brightness between a
given frame and the reference frame. This method was applied separately for
every SSO subfield, that is SSO-1, SSO-2${}+{}$SAAO, and SSO-3.

Our instrumental $V$ magnitudes and $(V-I_{\rm C})$ colors were transformed to standard ones using 
the updated Stetson's (2000) photometric database for M\,79. 
We obtained the following transformation 
equations:

\vskip0.6\baselineskip plus1pt minus1pt
 \hbox to\hsize{\hfill\vbox{\tabskip=1pt 
  \halign{%
   \hfil#\tabskip=5.5pt&%
   #\hfil\tabskip=0.7cm&%
   #\hfil\tabskip=1pt\cr
   \omit\span\omit SSO-1 $(N=56):$\hfil\cr
   $V-v$&         = $(0.067\pm0.009)\times(v-i)+(1.533\pm0.004)$,& $\sigma={}$0.019,\cr
   $V-I_{\rm C}$& = $(1.052\pm0.005)\times(v-i)+(0.499\pm0.002)$,& $\sigma={}$0.011,\cr
  }}\hfill}

\vskip0.4\baselineskip plus1pt minus1pt

 \hbox to\hsize{\hfill\vbox{\tabskip=1pt 
  \halign{%
   \hfil#\tabskip=5.5pt&%
   #\hfil\tabskip=0.7cm&%
   #\hfil\tabskip=1pt\cr
   \omit\span\omit SSO-2${}+{}$SAAO $(N=211):$\hfil\cr
   $V-v$&         = $(0.054\pm0.003)\times(v-i)+(1.451\pm0.001)$,& $\sigma={}$0.014,\cr
   $V-I_{\rm C}$& = $(1.046\pm0.003)\times(v-i)+(0.542\pm0.001)$,& $\sigma={}$0.012,\cr
  }}\hfill}

\vskip0.4\baselineskip plus1pt minus1pt

 \hbox to\hsize{\hfill\vbox{\tabskip=1pt 
  \halign{%
   \hfil#\tabskip=5.5pt&%
   #\hfil\tabskip=0.7cm&%
   #\hfil\tabskip=1pt\cr
   \omit\span\omit SSO-3 $(N=10):$\hfil\cr
   $V-v$&         = $(0.003\pm0.015)\times(v-i)+(1.517\pm0.012)$,& $\sigma={}$0.027,\cr
   $V-I_{\rm C}$& = $(0.975\pm0.009)\times(v-i)+(0.578\pm0.007)$,& $\sigma={}$0.015,\cr
  }}\hfill}
\vskip0.4\baselineskip plus1pt minus1pt

\noindent
where uppercase letters ($V$, $I_{\rm C}$) denote standard magnitudes
and lowercase letters ($v$, $i$), the instrumental magnitudes;
$N$ is the number of stars used in transformation and
$\sigma$ is the standard deviation of the fit.
It should be noticed that Stetson's (2000) photometry is tied directly to the Landolt's (1992)
standards.

Figure \figstddev\ shows the standard deviation for all stars with realiable photometry, 
that is stars located outside the cluster core, as a function of the mean $V$ magnitude. 
The cluster core is defined as a circular area around cluster center with a radius of 50 arcsec.
Our differential photometry has an accuracy of about 10 mmag for the brightest stars, decreasing to
40 mmag for stars with $V={}$20 mag.

\begin{figure}[tb]
\hbox to\hsize{\hss\includegraphics{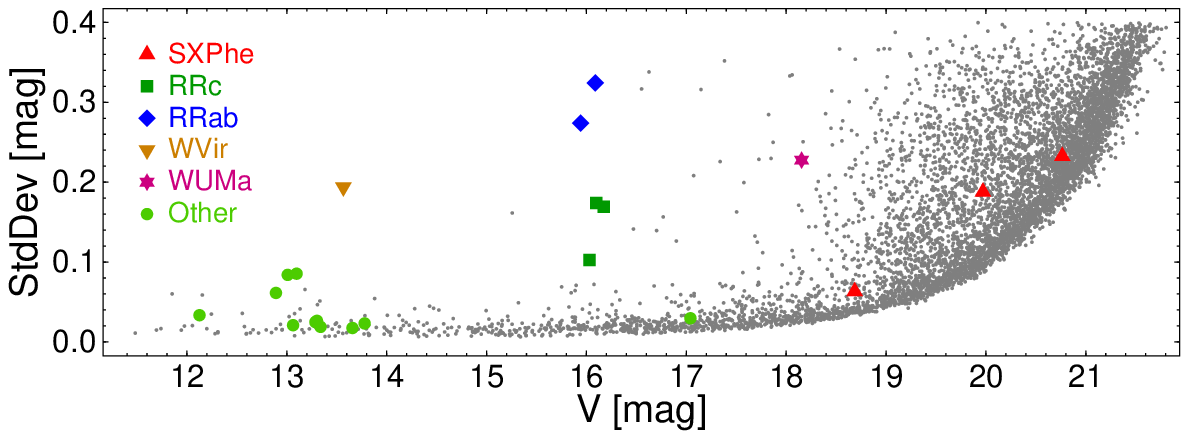}\hss}
\FigCap{%
 Standard deviation as a function of the average $V$ magnitude for light curves of 
 stars located outside cluster core (with distance from cluster's center $r>{}$50 arcsec).
 Variable stars are indicated with different black symbols.
 Note a group of RR Lyr stars at $V\approx{}$16.1 mag.}
\end{figure}
 
 In order to search for variable stars in M\,79, we reduced our CCD frames  
 with the image subtraction method (hereafter ISM) as implemented by Alard and Lupton (1998)
 and Alard (2000). 
 In general, the same procedure of reductions as in Kopacki
 (2000) was followed. 
 Only subfields containing cluster core, SAAO and SSO-2, were processed in this way.
 The two other subfields, SSO-1 and SSO-3 were searched for variable stars using DAOPHOT
 photometry. Initially, we reduced SAAO and SSO-2 frames separately, using reference
 frames constructed for each subfield from the best-seeing low-background images.
 We computed difference images and derived the final profile photometry from them using 
 two settings of the size of the convolution kernel: 18 px for the SSO frames and 12 px for the
 SAAO frames. These values secured adequate image subtraction, especially for stars located 
 in the cluster core or in the neighborhood of bright stars.

\begin{figure}[tb]
\hbox to\hsize{\hss\includegraphics{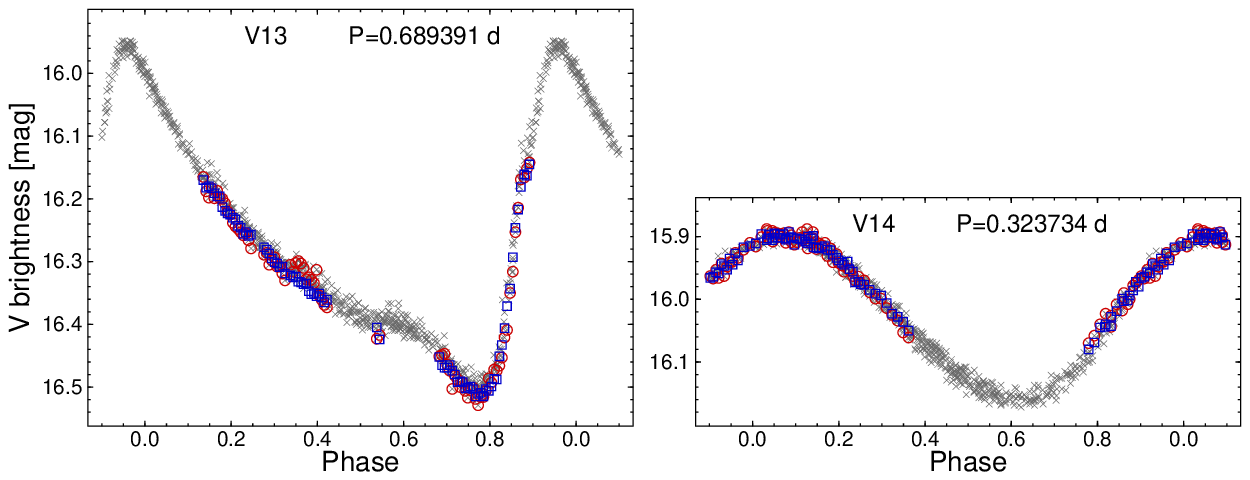}\hss}
\FigCap{%
 Comparison of the SAAO photometries for two RR Lyr stars, V13 ({\em left panel\/}) and
 V14 ({\em right panel\/}), obtained using SAAO (blue squares) and SSO-2 (red circles) reference frames. 
 Light gray crosses indicate the complete light curves covering both SAAO and SSO-2 observations, which
 were determined using SSO-2 reference frame. The ordinate scale is the same in both panels.}
\end{figure}

 To have ISM light curve of a given variable star expressed in the uniform flux scale, the
 differential photometry obtained from reductions of the SAAO frames needs
 to be transformed to the flux scale of the SSO-2 frames, for example.
 This task is quite simple for strictly periodic stars, even if -- as in our case --
 the two sets of images do not overlap in time.
 For stars showing long-term (possibly quasi-periodic) light-curve modulation it
 can be, however, troublesome and rather unreliable. In order to solve this problem, we
 decided to reduce
 our SAAO frames using the SSO-2 reference frame.
 The procedure that followed was very similiar to the original reductions with
 the ISM, namely all SAAO frames were interpolated onto the coordinate
 system of the SSO-2 reference frame, and then the resampled images were
 optimally convolved with the reference frame. Finally, the convolved frames were subtracted from
 reference frame, providing difference images. Since SAAO frames were taken 
 only through the $V$ filter, these calculations were performed only for this passband.
 
 It turned out, however, that the build-in application for image resampling
 is strictly valid only for translated frames (shifted along one or both rims). 
 It uses two-dimensional spline interpolator.
 Due to its simplicity, it works quite fast, 
 but cannot be used for frames rotated in respect to the reference frame 
 with an angle even of 0.2 degrees, especially if the image is large.
 In order to override this limitation, we wrote a programme {\tt ftsresample},%
 \footnote{http://www.astro.uni.wroc.pl/ludzie/kopacki/ftsresample.tar.gz}
 which also utilizes the spline method but in a local regime, in which resampling of each pixel
 in the original image is done separately using a box of nearby pixels.
 The method works properly for every rotation (and coordinate's shifting and scaling) but requires much
 computational time. Moreover, it has an advantage of transforming
 frames with different sizes into the same astrometric system.
 
 Comparing the light curves determined from the SAAO data in the approach described above 
 with those obtained from initial reductions, that is using two different reference frames,
 the SSO-2 and SAAO, we found them in excellent agreement. This testifies
 that the new method works well. Examples of the comparison
 for two RR Lyr stars are shown in Fig.\ \figismcomp.  
 The only possible source of systematic errors could be
 different realizations of a given passband, but evidently in this case they were negligible.
 
 We used two methods to search for variable stars in the set of $V$-filter difference
 images produced with the ISM for merged SSO-2 and SAAO subfields. Both methods were based on determination
 of light curves for fixed positions in the reference frame and a search for 
 periodic signal among them using Fourier analysis.
 In the first method, we derived differential fluxes for all stars detected in the reference
 frame by DAOPHOT package and for all brightenings found in the variability map
 (defined as an average of the absolute values of the best-seeing difference frames) and
 computed Fourier spectra of these light curves in the frequency range from 0 to
 50 d$^{-1}$. In order to identify variable stars, we inspected visually all stars showing a
 significant peak (with signal-to-noise ratio S/N${}>{}$4) in the periodograms of their light curves. 
 The results of this approach are summarized in Fig.\ \figmaxamp.
 This figure shows the plot of the S/N of the hightest peak in the amplitude spectrum as 
 a function of the corresponding frequency for all stars detected in the reference frame.
 In this way, we detected 33 variable stars, among which 20 are new discoveries.
 
 To check if we did not miss faint large-amplitude variable stars (especially in the crowded cluster
 core) we also performed the so called pixel photometry on difference frames. This approach
 assumes that a star is centered at a given pixel and the light curves are derived
 for all positions covering the whole area of the detector with a 1 px resolution. 
 The method was successfully applied by Kopacki (2005, 2007) to search for SX Phe 
 stars in the cores of M\,13 and M\,92. We confirmed the already obtained results but 
 found no other variable stars in the field.

\begin{figure}[tb]
\hbox to\hsize{\hss\includegraphics{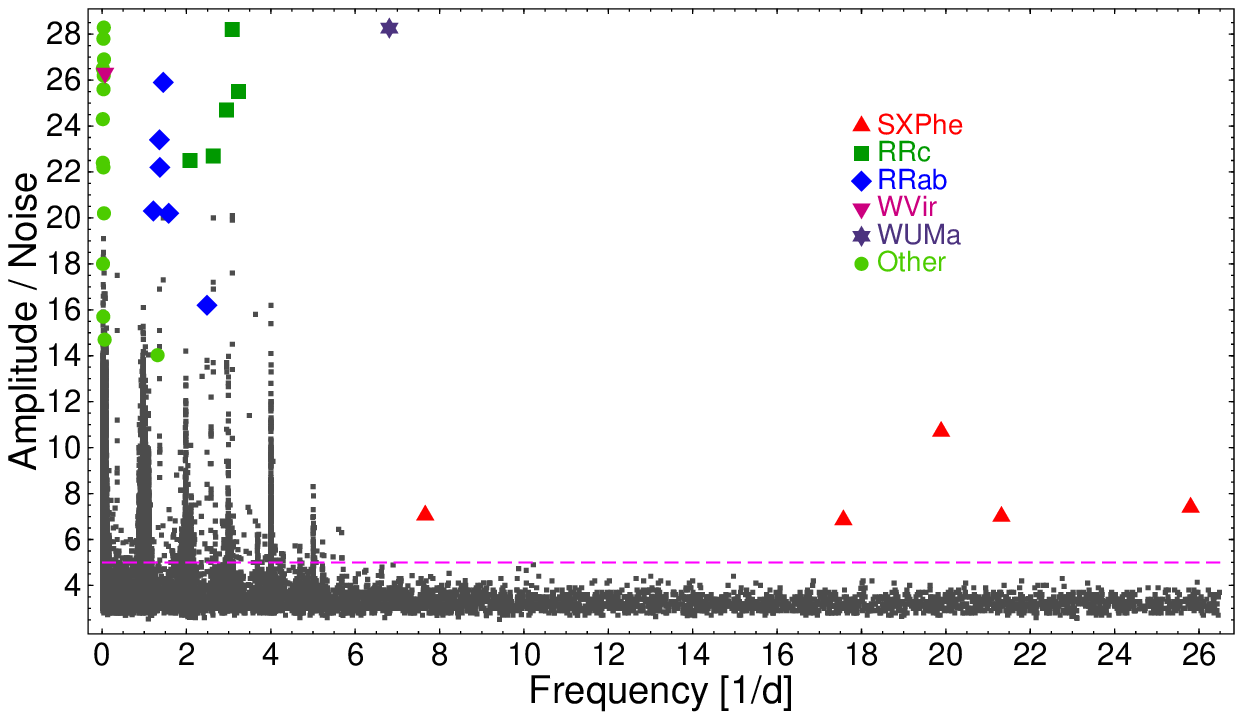}\hss}
\FigCap{%
 The amplitude-to-noise ratio 
 of the highest peak in the power spectrum plotted against the peak's
 frequency. Each point corresponds to one star detected in the
 reference frame. Different types of variable stars are indicated
 with various symbols.}
\end{figure}

 With the purpose of deriving standard $V$-magnitudes for pulsating variable stars located
 in the cluster crowded core, for which our ground-based DAOPHOT photometry is
 not realiable, we made also use of the available HST observations of M\,79.
 Among the archival WFPC2 HST data we found two data-sets. The first set 
 consists of three F439W (equivalent to $B$) and two F555W ($V$) frames, which were
 taken on 1995 Oct 14 within the HST proposal GO\,6095 by G.\ Djorgovski. The other
 observations are four F336W ($U$) and five F555W ($V$) images obtained
 on 1997 Apr 18 within proposal GO\,6607 by F.\ Ferraro. All these data were used
 by Lanzoni {\em et al.\/} (2007) in their panchromatic study of M\,79.
 In the context of our variability
 survey, the HST data we used consist virtually of only two epochs.
 
 The outlines of the WFPC2 fields are 
 shown in Fig.\ \figfldcore. As can be seen, they were chosen in such a way, that
 cluster core is approximately centered on the PC chip (giving the highest resolution
 of 0.05 arcsec/px). Moreover, they embrace the area 
 where variables V3 -- V5, V9 -- V14, V16, and V18 are located. 
 We downloaded the WFPC2 data from the Multimission Archive at Space Telescope%
 \footnote{http://archive.stsci.edu}
 and processed them with the HSTphot package of Dolphin (2000a, 2000b). 
 We followed the standard procedure of reductions as described in the manual to
 this package. 
 
 Using standard $V$ magnitudes obtained from the HST frames we were able to
 transform our differential $V$-flux photometry ($\Delta f$) from the ISM to magnitude scale
 by fitting for every variable star the following equation:
 $$ V - V_{\rm ref} = -2.5\log(1+\Delta f/f_{\rm ref}),$$
 where $V_{\rm ref}$ and $f_{\rm ref}$ are reference points in magnitude and flux, respectively, the two
 parameters which are to be found. Since the HST photometry and our observations are separated by about 12 yr,
 the fit was performed in the
 phase domain, rather than in the time domain. For RR Lyr stars we assumed there is no phase 
 shift between satellite and ground observations.
 For SX Phe stars (having much shorter periods than RR Lyr stars) we had to apply small phase
 shift to properly align the ISM and HST light curves, but always the fit was unambigous.
 In this manner, $V$ light curves were determined for two SX Phe stars, V16 and V17, and
 for five RR Lyr stars, V5, V9, V10, V12, and V13.

\begin{figure}[tb]
\hbox to\hsize{\hss\includegraphics{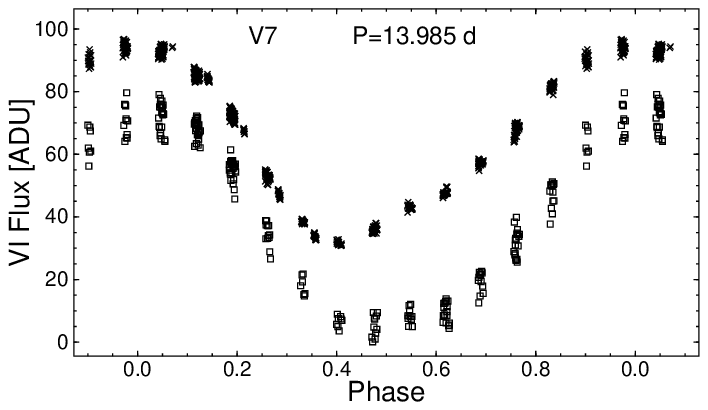}\hss}
\FigCap{%
 $V$-filter (crosses) and $I_{\rm C}$-filter (squares) light curves of a W Vir star V7.}
\end{figure}
 
\section{Results of the Variability Survey}

 All fourteen variable or suspected variable stars listed in the CVSGC for M\,79 
 were inside the field we observed. Star V1 was found to be constant in our data.
 We confirm that V7 is a Pop.\ II Cepheid of the W Vir type. For the first time, we determine 
 the period of this star which is equal to 13.985 d and obtain $VI_{\rm C}$ light curves 
 very well covered in phase. They are shown in Fig.\ \figlcwvir. Two
 stars, V2 and V8, are semiregular variables located at the tip of
 the red giant branch in the cluster color -- magnitude diagram (see Fig.\ \figcmd). 
 It should be noted that we were not able to derive light curves expressed in magnitudes for
 all observed variable stars. To keep the presentation of our results consistent as much as
 possible, light curves of variables from SSO-2 subfield will be shown in flux units, whereas
 light changes of variables from other two subfields, with standard magnitudes.

\begin{figure}[tb]
\hbox to\hsize{\hss\includegraphics{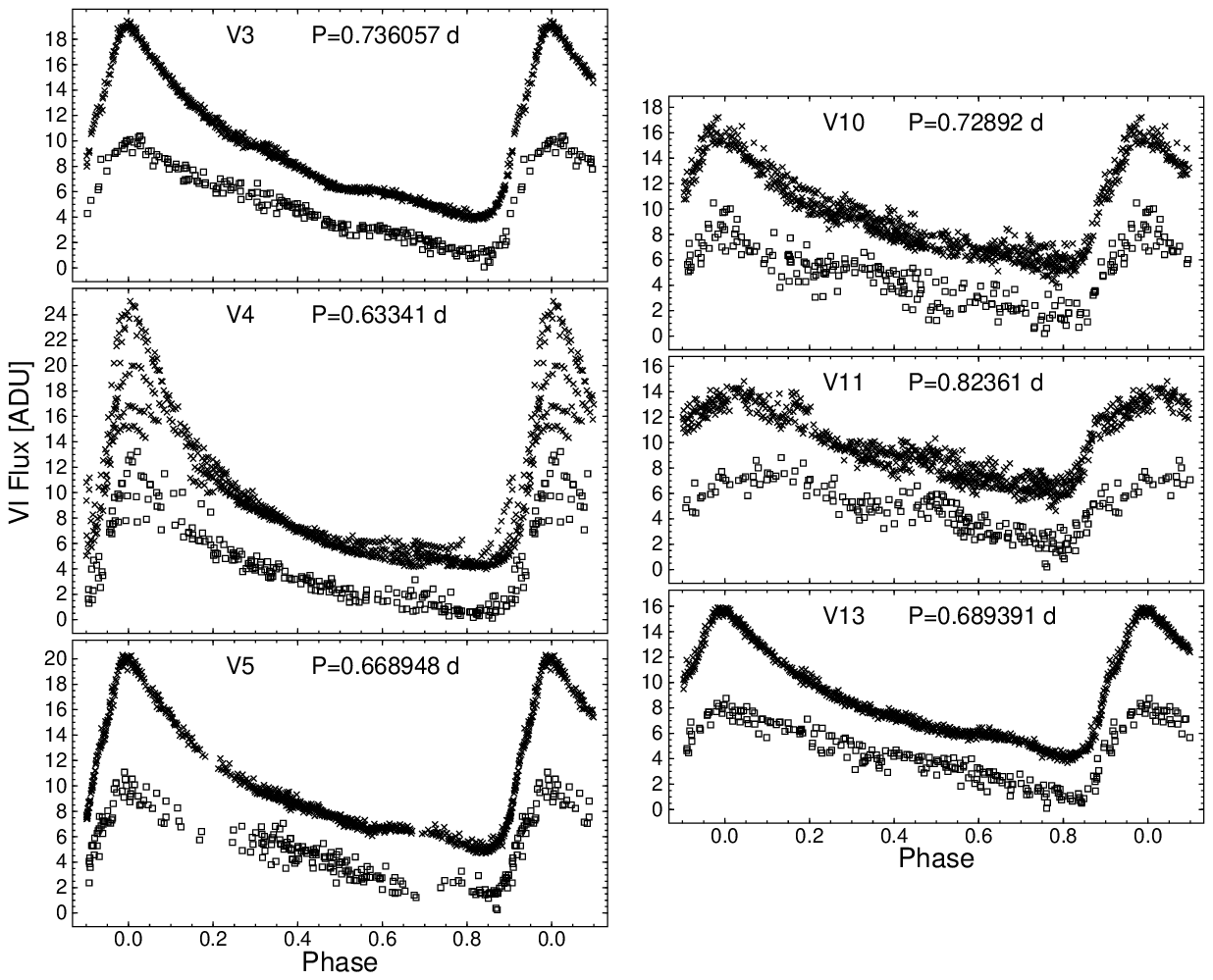}\hss}
\FigCap{%
 $V$-filter (crosses) and $I_{\rm C}$-filter (squares) light curves of RRab stars in M\,79.
 The ordinate scale is the same in all panels.}
\end{figure}
 
\begin{figure}[tb]
\hbox to\hsize{\hss\includegraphics{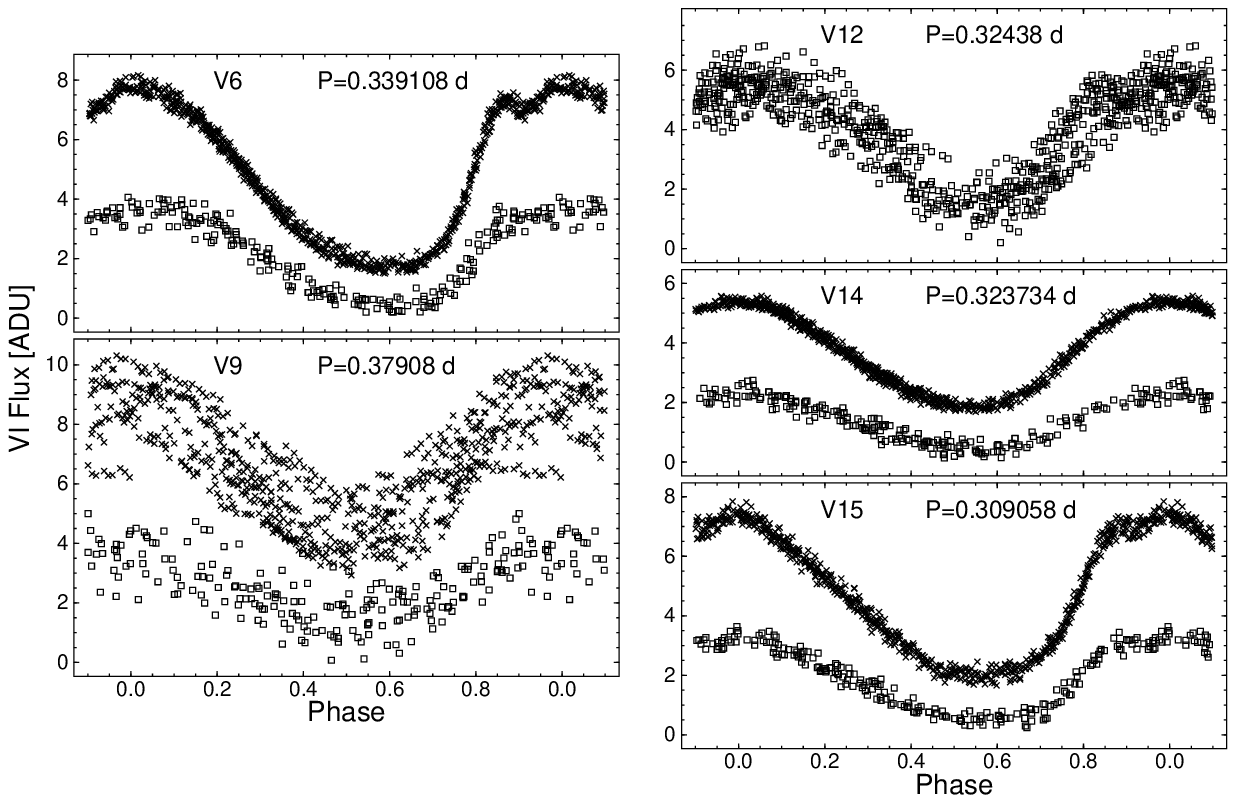}\hss}
\FigCap{%
 $V$-filter (crosses) and $I_{\rm C}$-filter (squares) light curves of RRc stars in M\,79. Only V15
 was discovered to be variable in this work. The ordinate scale is the same in all panels.}
\end{figure}
 
 We observed all ten RR Lyr stars previously known in M\,79 and found one
 more star of this type, surprisingly at a very large angular distance from the cluster
 centre, 11.4 arcmin (in Fig.\ \figfld\ it can be found close to the western border of the
 SSO-2 field). It is an RRc star. Its position at the horizontal branch of
 the cluster's color -- magnitude diagram (Fig.\ \figcmd) cleary indicates it is a cluster member.
 Extending the numbering scheme
 of the CVSGC for M\,79, we denote this new variable star V15.
 The $V$- and $I_{\rm C}$-filter light curves of all RRab and RRc stars we observed are shown in 
 Figs.\ \figlcrrab\ and \figlcrrc, respectively. Since all RR Lyr stars in a given globular cluster
 have approximately the same mean brightness, differences in flux amplitude of the observed RR Lyr variables
 strictly reproduce differences in amplitude expressed in magnitude.
 The larger scatter of observations
 in the light curves of V10, V11, and V12 is very likely of instrumental origin and is caused by
 imperfections of image subtraction in the cluster core. However, one of the RRc stars, V9, clearly shows 
 a modulation of the light curve. 
 This star will be discussed later.
 
 In addition, we have found seven other periodic variable stars. Three of them were detected in the central
 SSO-2 subfield. We designate them V16 through V18 and classify as SX Phe stars. Their location in the
 blue straggler region of the cluster's color -- magnitude diagrams (see the left panel of Fig.\ \figcmd\ and Fig.\ \figcmdhst) 
 strongly suggests they are physically bound to the cluster.
 The $V$-filter light 
 curves of these SX Phe stars are presented in Fig.\ \figlcsxphe. 
 All these stars, 
 except V18, exhibit single-period oscillations. Multiperiodic variable V18 will be discussed
 separately. 

 The remaining four periodic variable stars, which we name V29 through V32, were found in the side subfield SSO-3,
 and can be identified in Fig.\ \figfld. Judging from their location in the color -- magnitude
 diagram (the right panel of Fig.\ \figcmd) they are likely field stars.
 The light curves of these variables are shown in Fig.\ \figlcother.
 Variables V29 and V30 exhibit changes with short periods typical for SX Phe and $\delta$ Sct stars.
 The light curve of V29 is clearly asymmetric with a sharp maximum, which indicates for a 
 high-amplitude $\delta$~Sct or SX Phe star. Star V30 can be either a pulsating star or a
 W~UMa type eclipsing star or an ellipsoidal system. If it is a binary system, its period should be doubled. 
 Finally, V31 is a W UMa type variable and V32 shows brightness
 variations of unknown type with a period of 0.7587 d. 
 
\begin{figure}[tb]
\hbox to\hsize{\hss\includegraphics{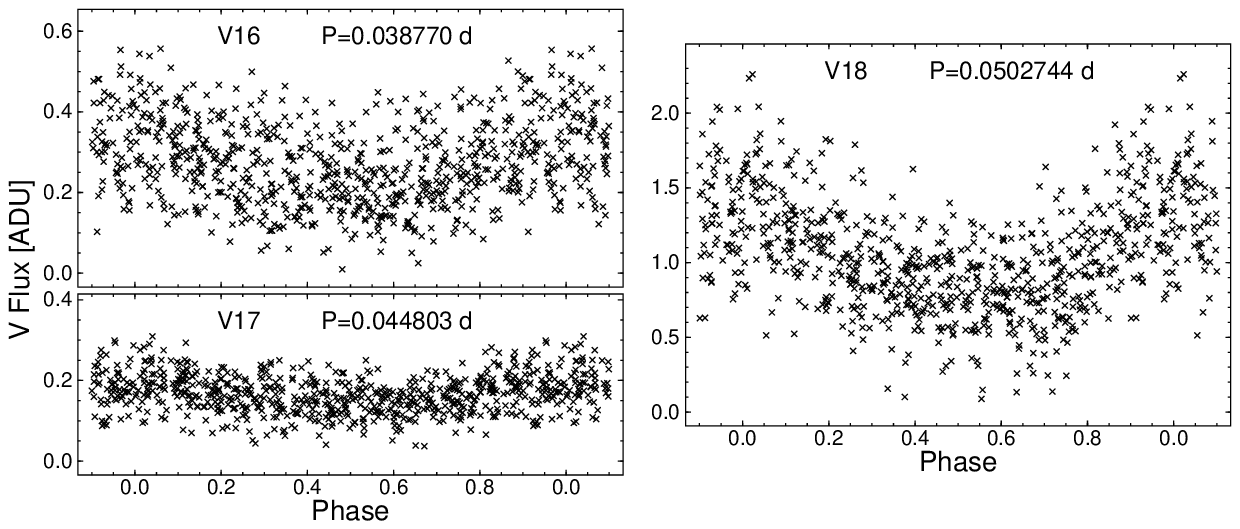}\hss}
\FigCap{%
 $V$-filter light curves of SX Phe stars in M\,79. They all were
 discovered in this work.}
\end{figure}

\begin{figure}[tb]
\hbox to\hsize{\hss\includegraphics{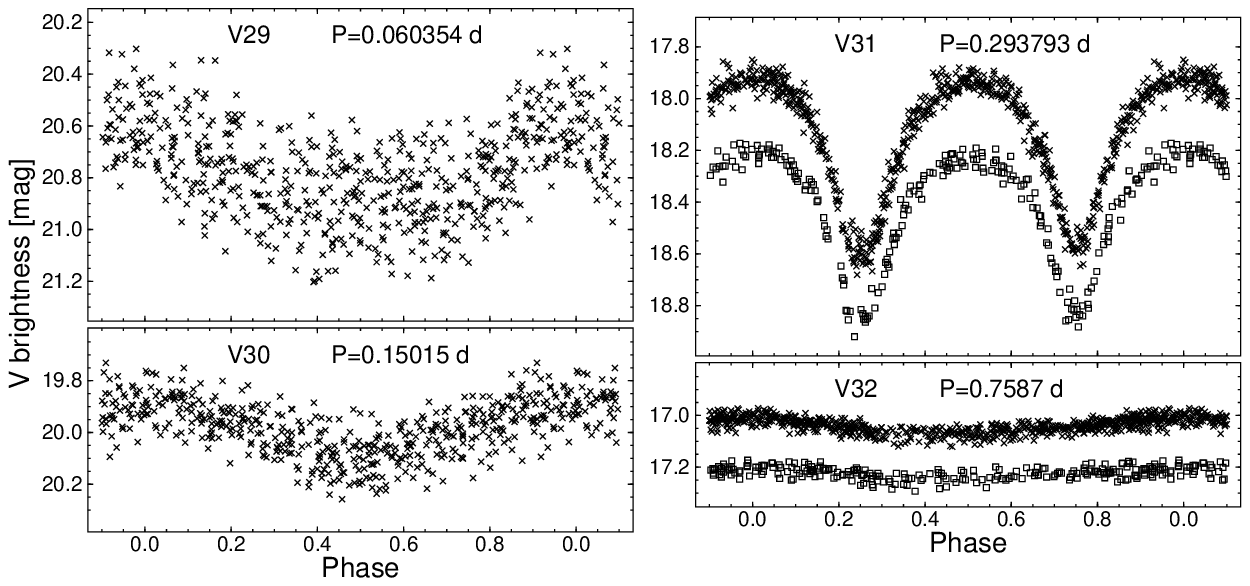}\hss}
\FigCap{%
 $V$-filter (crosses) and $I_{\rm C}$-filter (squares) light curves of other short-period variable stars 
 detected in the SSO-3 field of M\,79. All these stars are new discoveries. The ordinate scale is the 
 same in all panels.}
\end{figure}

 Bright red giants in globular clusters usually show small-amplitude irregular 
 variability (Welty 1985). In M\,53 and M\,13, for example, almost all stars at 
 the tip of the red giant branch are variable (Kopacki 2000, 2003). In contrast, no or very few variable red
 giants were found in M\,92 (Kopacki 2001) and M\,80 (Kopacki 2013). From our observations of M\,79 we find
 almost all red giants at the very tip of red giant branch to be variable. They mostly
 exhibit small-amplitude long-period/irregular light changes. Two of them, V2 and V8, were previously
 known as variable, and the other twelve (including three field stars), which we designate, 
 V19 through V28, V33, and V34, were discovered in this study. 
 Those variable red giants which have realiable DAOPHOT photometry are indicated in the color -- magnitude 
 diagram of the cluster (left panel of Fig.\ \figcmd).

\begin{table}[htb]
 \TabCap{12.5cm}{Astrometric and photometric data for variable stars in the field of M\,79}
 \TableFont
 \def\hst{\hbox to0pt{$^\star$\hss}} 
 \def\bn{\setbox0=\hbox{0}\hskip\wd0}
 \def\mp{\omit\hfil--\hfil}
 \hbox to\hsize{\hfill\vbox{\tabskip=2pt 
  \halign{%
   #\hfil\tabskip=9pt&
   \hfil#\hfil&
   \hfil#\hfil&
   \hfil#\hfil&
   \hfil#\hfil&
   \hfil#\hfil&
   \hfil#\hfil&
   #\hfil&
   #\hfil\tabskip=2pt\cr 
   \noalign{\hrule\vskip3pt}
   Var&
   $\alpha_{2000}$& 
   $\delta_{2000}$& 
   $\langle V\rangle$& 
   $\langle V-I_{\rm C}\rangle$& 
   \omit\hfil$\Delta V$\hfil& 
   \omit\hfil$\Delta I_{\rm C}$\hfil& 
   \omit\hfil$P$\hfil&
   Type\cr
   \noalign{\vskip1pt}
     & [$^{\rm h}$ $^{\rm m}$ $^{\rm s}$]& [$^\circ$ $^\prime$ $^{\prime\prime}$]&
     [mag]& [mag]& [mag]& \omit\hfil[mag]\hfil& \omit\hfil[d]\hfil&\cr
   \noalign{\vskip3pt\hrule\vskip3pt}
  V1& 5 24 13.00& $-$24 34 43.8& 15.020&  1.072&    --&    --&       --&   const\cr
  V2& 5 24 16.74& $-$24 32 34.3& 13.092&  1.428& 0.220& 0.159&     31.3&      VRG\cr
  V3& 5 24 13.54& $-$24 32 29.1& 15.911&  0.542& 0.888& 0.550& \bn{}0.736057& RR Lyr\cr
  V4& 5 24 17.77& $-$24 32 16.2& 16.036&  0.504& 0.999& 0.734& \bn{}0.63341&  RR Lyr\cr
  V5\hst& 5 24 10.23& $-$24 31 03.6& 16.059&     --& 0.606&    --& \bn{}0.668948& RR Lyr\cr
  V6& 5 24 06.03& $-$24 29 32.9& 16.106&  0.363& 0.481& 0.284& \bn{}0.339108& RR Lyr\cr
  V7& 5 24 12.68& $-$24 31 41.9& 13.644&  0.903& 0.516& 0.392&     13.985&    W Vir\cr
  V8& 5 24 11.54& $-$24 31 38.2&     --&     --&    --&    --&        \mp&    VRG\cr
  V9\hst& 5 24 12.58& $-$24 31 52.6& 16.196&     --& 0.423&    --& \bn{}0.37908&  RR Lyr\cr
      &&&&&                                      0.297&    --& \bn{}0.36099&  \cr 
      &&&&&                                      0.108&    --& \bn{}0.37037&  \cr 
 V10\hst& 5 24 12.13& $-$24 31 34.5& 16.061&     --& 0.781&    --& \bn{}0.72892&  RR Lyr\cr
 V11& 5 24 11.93& $-$24 31 34.6&     --&     --&    --&    --& \bn{}0.82361&  RR Lyr\cr
 V12\hst& 5 24 11.35& $-$24 31 28.3& 16.220&     --& 0.531&    --& \bn{}0.32438&  RR Lyr\cr
 V13\hst& 5 24 10.59& $-$24 31 11.5& 16.265&     --& 0.548&    --& \bn{}0.689391& RR Lyr\cr
 V14& 5 24 07.77& $-$24 31 00.3& 16.025&  0.364& 0.258& 0.145& \bn{}0.323734& RR Lyr\cr
 V15& 5 23 23.74& $-$24 27 46.3& 16.173&  0.319& 0.465& 0.283& \bn{}0.309058& RR Lyr\cr
  \noalign{\vskip3pt}
 V16\hst& 5 24 09.97& $-$24 31 07.3& 18.580&     --& 0.433&    --& \bn{}0.038770&  SX Phe\cr
 V17& 5 24 14.15& $-$24 33 20.7& 18.687&  0.443& 0.027& 0.016& \bn{}0.044803&  SX Phe\cr
 V18\hst& 5 24 10.86& $-$24 31 11.8& 18.255&     --& 0.833&    --& \bn{}0.0502744& SX Phe\cr
      &&&&&                                      0.349&    --& \bn{}0.0391700& \cr 
  \noalign{\vskip3pt}
 V19& 5 24 12.26& $-$24 31 25.2&     --&     --&    --&    --&     \mp& VRG\cr
 V20& 5 23 20.18& $-$24 28 03.2& 13.008&  1.543& 0.243& 0.158&     \mp& VRG\cr
 V21& 5 24 09.01& $-$24 31 15.4&     --&     --&    --&    --&    29.1& VRG\cr
 V22& 5 24 09.57& $-$24 29 51.3& 13.337&  1.439& 0.028& 0.039&     \mp& VRG\cr
 V23& 5 24 09.61& $-$24 30 26.5& 13.216&  1.462& 0.030& 0.044&     \mp& VRG\cr
 V24& 5 23 47.33& $-$24 30 37.5& 13.310&  1.454& 0.074& 0.048&     \mp& VRG\cr
 V25& 5 24 40.48& $-$24 29 00.0& 13.783&  1.024& 0.055& 0.048&    21.0& Irr\cr
 V26& 5 24 18.48& $-$24 31 43.2& 13.289&  1.494& 0.053& 0.040&     \mp& VRG\cr
 V27& 5 24 09.13& $-$24 34 15.9& 13.059&  1.503& 0.032& 0.045&     \mp& VRG\cr
 V28& 5 24 18.04& $-$24 30 14.7& 13.662&  1.289& 0.028& 0.038&     \mp& VRG\cr
  \noalign{\vskip3pt}
 V29& 5 24 21.93& $-$24 39 03.8& 20.774&  0.499& 0.297& 0.231& \bn{}0.060354& SX Phe\cr
 V30& 5 24 00.02& $-$24 38 37.6& 19.985&  0.704& 0.205& 0.173& \bn{}0.15015&  SX Phe\cr
 V31& 5 24 57.58& $-$24 43 17.9& 18.129&  0.820& 0.690& 0.662& \bn{}0.293793& W UMa\cr
 V32& 5 24 18.92& $-$24 49 39.2& 17.045&  1.280& 0.064& 0.045& \bn{}0.7587&   --\cr
  \noalign{\vskip3pt}
 V33& 5 24 31.01& $-$24 21 38.4& 12.113&  0.857& 0.089& 0.079& 23.09& Irr\cr
 V34& 5 24 10.86& $-$24 13 40.1& 12.857&  2.620& 0.183& 0.063& 45.3&  Irr\cr
  \noalign{\vskip3pt\hrule}
  }}\hfill}
\end{table}

\begin{figure}[tb]
\hbox to\hsize{\hss\includegraphics{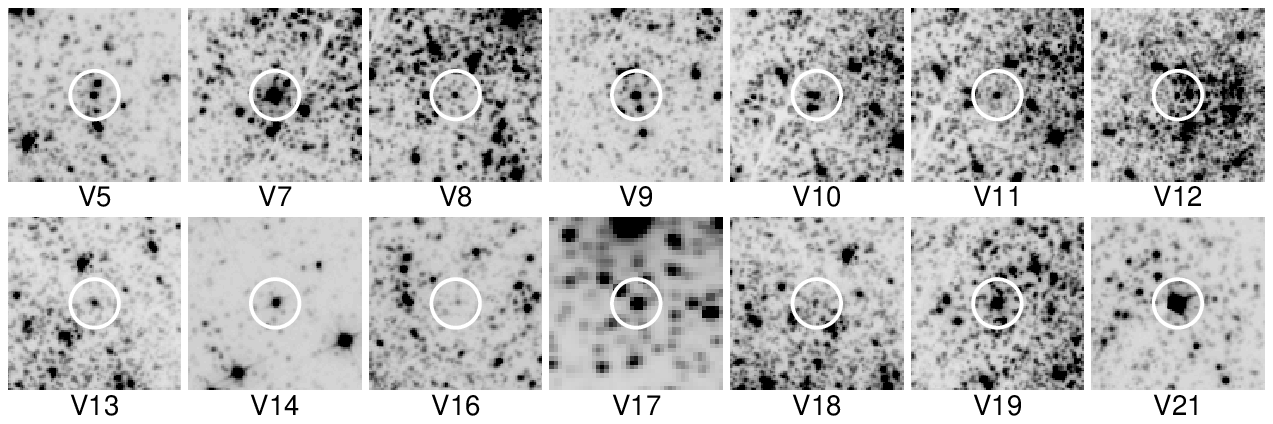}\hss}
\FigCap{%
 Finding charts for variable stars located in the central region of M\,79. Except for V17, 
 they were extracted from the WFPC2 frames and cover 15${\times}{}$15 arcsec$^2$.
 The chart for v17 was prepared using SSO-2 reference frame and its size is 
 30${\times}{}$30 arcsec$^2$. North is up and east to the left.}
\end{figure}

 All variable stars observed in and around M\,79 are indicated in Figs.\ \figfld\
 and \figfldcore. Fig.\ \figfldcore\ shows the dense central area of the
 cluster. 
 In addition, finding charts
 for variables located in the very central region of the cluster 
 ($r<{}$1 arcmin) and for V17 are presented in Fig.\ \figvarcharts.
 The astrometric and photometric parameters of the variable stars 
 are given in Table \tabvarpar.
 For each star we provide adopted designation, equatorial coordinates,
 $(\alpha,\delta)_{2000}$, and if available, average $V$ magnitude, $\langle V\rangle$, 
 mean $(V-I_{\rm C})$ color index, $\langle V-I_{\rm C}\rangle$,
 ranges of variability, $\Delta V$ and 
 $\Delta I_{\rm C}$,
 period(s), $P$, and type of variability. The equatorial coordinates were
 derived from astrometric transformation of our rectangular positions using about
 750 (SSO-1), 2680 (SSO-2) and 540 (SSO-3) stars in common with the NOMAD catalogue (Zacharias {\em et al.\/}\ 2004).
 Depending on the SSO subfield, the standard deviations of the derived astrometric equations were 
 in the range 0.2${}-{}$0.35 arcsec for right ascention and declination.
 The mean magnitudes $\langle V\rangle$ and $\langle I_{\rm C}\rangle$ are intensity-weighted averages in the case of pulsating stars and
 arithmetic means for the other variable stars. The average colors are defined as $\langle V\rangle-\langle I_{\rm C}\rangle$.
 The intensity-weighted mean magnitudes and the ranges of variability of periodic stars were computed from 
 the fit of the truncated Fourier series to our instrumental $V$ and $I_{\rm C}$ magnitudes and only 
 for stars having reliable photometry. 
 $\Delta V$ for V4 should be treated with caution, since this RRab star shows strong Blazhko effect
 (Fig.\ \figlcrrab). 
 For the variable red giants (VRGs) we give the range of the
 mean nightly magnitudes. Periods are given with 
 an accuracy resulting from a non-linear least-squares fit of truncated Fourier
 series to the observations. For each star we used the
 smallest possible number of harmonic components which described reasonably well the observed
 light curve. For multiperiodic stars, V9 and V18, we provide 
 ranges of variability and periods of all detected terms.
 The stars for which photometric parameters were derived using HST observations are indicated in
 Tab.\ \tabvarpar\ with an asterisk symbol following star designation.

 Altogether, the $V$- and $I_{\rm C}$-filter light curves expressed in flux units have been 
 obtained for 33 variable stars. They can be downloaded from the {\em Acta
 Astronomica Archive}. Moreover, for stars having reliable DAOPHOT or HST photometry we provide
 light curves with differential fluxes transformed to magnitudes and then
 to the standard system. 

\begin{figure}[tb]
\hbox to\hsize{\hss\includegraphics{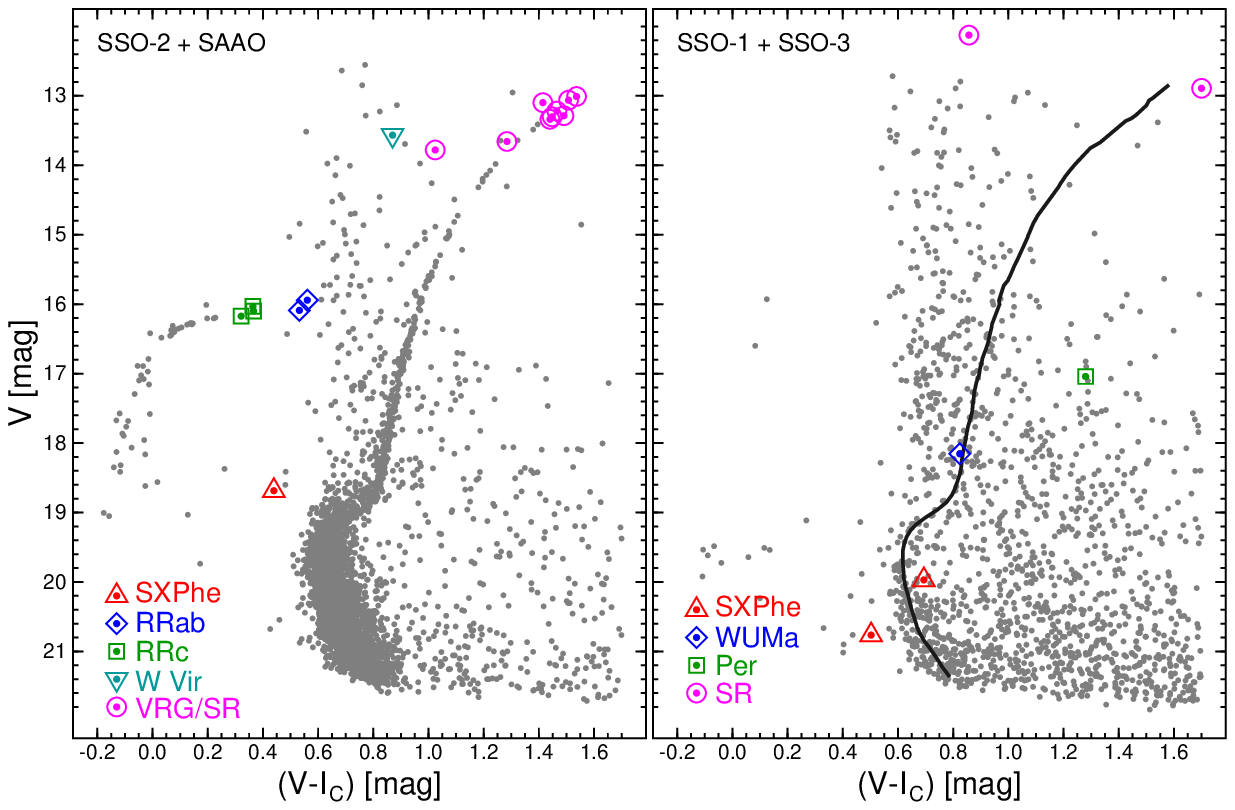}\hss}
\FigCap{%
 The $V$ {\em vs.\/}\ $V-I_{\rm C}$ colour -- magnitude diagrams for the combined SSO-2 and SAAO
 fields covering cluster core ({\em left panel\/}) and the peripherial SSO-1 and SSO-3 fields
 ({\em right panel\/}).
 In the {\em left panel\/} SX Phe stars are indicated with top-tipped triangles, RRc variables 
 with open squares, RRab stars with diamonds, W Vir star with a bottom-tipped triangle, and
 variable red giants with open circles. In the {\em right panel\/} 
 the field SX Phe stars are represented with triangles, the eclipsing system of the
 W UMa type with a diamond, periodic star of unknown type with open 
 square, and two semiregular variables with open circles. As a reference we
 show with solid line a fiducial sequence obtained from 
 the main sequence and red giant branch of the {\em left panel}.}
\end{figure}

\section{Color -- Magnitude Diagrams}

 The $V$ {\em vs.\/}\ $(V-I_{\rm C})$ diagrams for the fields observed
 in M\,79 are shown in Fig.\ \figcmd. 
 Only stars with the angular distance from cluster center larger than 1.5 arcmin and
 error in color smaller than 0.05 mag are taken into account. The left panel of this figure presents
 color -- magnitude diagram for the combined SSO-2 and SAAO fields, embracing 
 the core of the cluster, while in the right panel we show a diagram for peripherial
 fields SSO-1 and SSO-3. As expected, the SSO-1 and SSO-3 frames are heavily
 dominated with stars not belonging to the cluster. The W Vir star, V7, is located
 about 2.5 mag above cluster's horizontal branch, and
 SX Phe star, V17, is found in the blue straggler region. Only five RR Lyr stars
 could be included in Fig.\ \figcmd, two RRab stars, V3 and V4, and three RRc variables
 V6, V14 and V15. RRab stars clearly separate from RRc variables,
 which occupy the hotter part of the instability strip. 
 
 Fig.\ \figcmdhst\ presents colour -- magnitude diagrams obtained from the HST observations.
 In the left panel we show $V$ {\em vs.\/} $(B-V)$ diagram and in the right panel, $V$ {\em vs.\/} $U-V$ diagram.
 In these diagrams we indicate those variable stars, which could not be resolved in our ground-based data
 (and therefore are not shown in Fig.\ \figcmd), that is two SX Phe stars, V16 and V18, 
 and seven RR Lyr stars, V5, V9, and V10 -- V13.
 Whenever possible, we used cycle-averaged $V$ magnitudes derived from the SSO light-curves, but still all 
 colors were determined from sparce HST observations in $B$ and $U$. This is the reason for
 some RR Lyr stars to deviate from their expected position in the horizontal branch of the cluster.
 The most severe examples are two RRab stars, V3 (the brightest and bluest RR Lyr star
 in the left panel of Fig.\ \figcmdhst) and V10 (the bluest RR Lyr star in the right panel of Fig.\ \figcmdhst). 
 This is a consequence of the large amplitudes and non-sinusoidal shapes
 of the light curves for this type of pulsating stars. 
 The two SX Phe stars are found among the blue straggler population of the cluster.

\begin{figure}[tb]
\hbox to\hsize{\hss\includegraphics{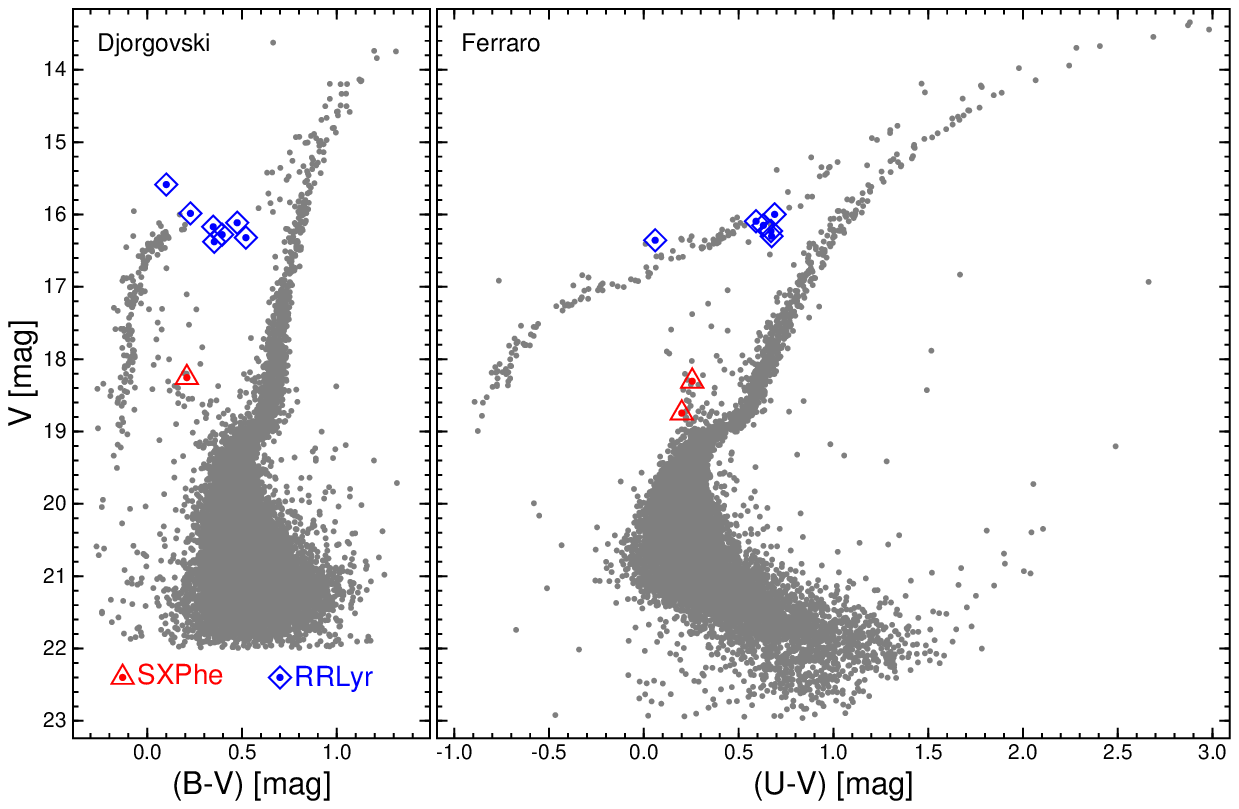}\hss}
\FigCap{%
 The $V$ {\em vs.\/}\ $(B-V)$ ({\em left panel\/}) and $V$ {\em vs.\/}\ $(U-V)$ 
 ({\em right panel\/}) colour -- magnitude diagrams from the HST WFPC2 observations.
 SX Phe stars are indicated with top-tipped triangles, while RR Lyr variables, 
 with open diamonds.}
\end{figure}

\begin{figure}[tb]
\hbox to\hsize{\hss\includegraphics{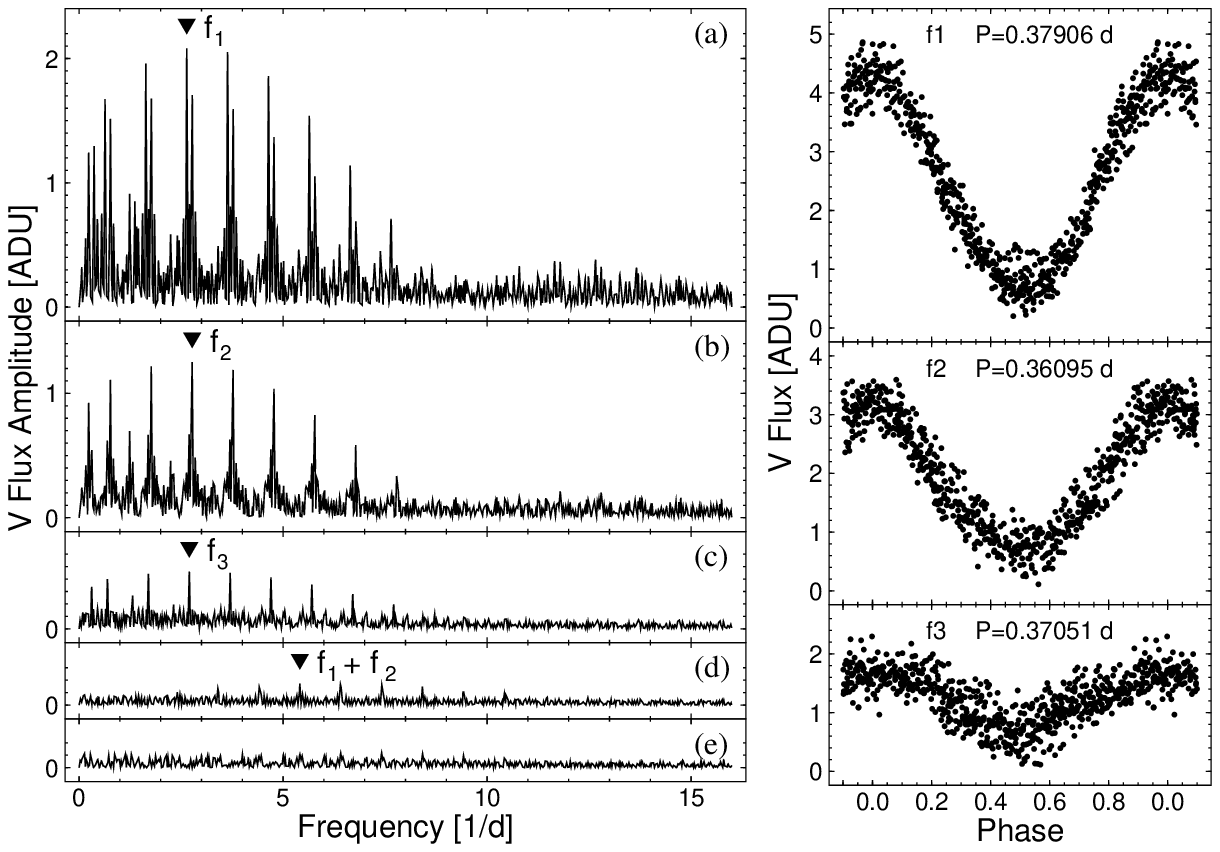}\hss}
\FigCap{%
 {\em Left panels:} Amplitude spectra of the RR Lyr star V9: 
 (a) for original $V$-filter observations,
 (b) after prewhitening with frequency $f_1={}$2.6380 d$^{-1}$,
 (c) after removing frequencies $f_1$ and $f_2={}$2.7702 d$^{-1}$,
 (d) after prewhitening with frequencies $f_1$, $f_2$, and $f_3={}$2.7000 d$^{-1}$,
 (e) after removing frequencies $f_1$, $f_2$, $f_3$ and the combination frequency $f_1+f_2$.
 The ordinate scale is the same in all panels.
 {\em Right panels:} Separated light curves of V9 for the $f_1$, $f_2$, and $f_3$
 components (from {\em top\/} to {\em bottom\/}) phased with appropriate periods.
 Arbitrarily chosen initial epoch is the same for all light curves.
 The ordinate scale is the same in all panels.}
\end{figure}
 
\section{Multiperiodic Variable Stars}

 Two RR Lyrae stars, an RRab star V4 (Fig.\ \figlcrrab) and an RRc star V9
 (Fig.\ \figlcrrc) show long-period modulation of the light-curve shape, which 
 is reffered to as the Blazhko effect (see, e.g., Kolenberg {\em et al.\/}\ 2010, Benk\H{o} {\em et al.\/}\ 2010).
 This behaviour is an indication of multiperiodicity. We searched for additional
 periodic signals in the light curves of all observed pulsating stars and found 
 that SX Phe star V18 is also multiperiodic.
 The modulation period of V4 turned out to be much longer then the time-span
 of our observations, so we could not make meaningful Fourier analysis of its
 light curve. We shall now discuss the remaining two stars, V9 and V18, in more detail.
 It should be noted that Amigo {\em et al.\/} (2011) and Kains {\em et al.\/} (2012) suggested, 
 that V9 could be an RRd star.
 All frequencies cited below are given with an accuracy resulting from the fit of the
 finite Fourier series to the observations. 

 The Fourier spectrum of the original observations of RRc star V9 is shown
 in Fig.\ \figmultivix{}a. The highest peak occurs at a frequency of $f_1={}$2.6380 d$^{-1}$.
 Periodogram of the residuals obtained after prewhitening the original data with the
 frequency $f_1$ is presented in Fig.\ \figmultivix{}b. The highest peak is found
 at a freqeuncy of $f_2={}$2.7702 d$^{-1}$, which is quite close to the primary frequency $f_1$.
 The third significant frequency, $f_3={}$2.7000 d$^{-1}$, can be seen in Fig.\ 
 \figmultivix{}c, after removing
 $f_1$ and $f_2$ from the original data. In the data prewhitened with all three
 detected frequencies we found only combination term $f_1+f_2$ and its daily
 aliases, which can be seen in the periodogram of these observations 
 illustrated in Fig.\ \figmultivix{}d.
 No more periodic terms are
 found in the light curve of V9, which is testified by the flat periodogram
 obtained after prewhitening the data with all observed frequencies, including
 combination term (see Fig.\ \figmultivix{}e).
 Phase diagrams for all three periodic components found in V9 are shown in the right panels of Fig.\
 \figmultivix{}.

 Summarizing, we detect in light changes of V9 a close frequency triplet with secondary and tertiary
 frequencies located on the high-frequency side of $f_1$.
 The differences in frequency are the following:
 $\Delta f_{2-1} = f_2-f_1=0.0620$ d$^{-1}$ and
 $\Delta f_{3-1} = f_3-f_1=0.1322$ d$^{-1}$. As can be seen, frequencies
 $f_1$, $f_2$, and $f_3$ do not form an equidistant triplet structure.
 Moreover, periodicity with the largest amplitude, which can be attributed
 to the radial mode, does not correspond to the central component. Taking
 into account these arguments, we conclude that V9 does not exhibit classical
 Blazhko effect, that is (almost) periodic modulation of amplitude and phase
 of light variations, which in the Fourier domain manifests as an
 equidistant multiplet structure with the central frequency identified with
 the radial mode (Kov\'acs 2009, Benk\H{o} {\em et al.\/}\ 2010). 
 A very small separation of the $f_2$ and $f_3$ components
 from the presumably radial one ($f_1$) suggests that both $f_2$ and $f_3$ correspond to 
 nonradial modes of pulsation. 
 
 The other multiperiodic star we found is the SX Phe variable V18. The Fourier
 spectrum of the $V$-filter data of this star is presented in Fig.\ \figmultivxviii{}a.
 The highest peak is found at frequency $f_1={}$19.8909 d$^{-1}$. The asymmetric shape
 of this periodic component is manifested by the presence of the peak at
 2$f_1$. After removing the frequency
 $f_1$ with its harmonics from original observations, we obtain residuals for which the  
 periodogram is shown in Fig.\ \figmultivxviii{}b. Here the highest peak occurs at 
 frequency of $f_2={}$0.4478 d$^{-1}$. This signal can be traced to
 the contamination from a nearby RR Lyr star V13. 
 Variable V13 is located only
 3.7 arcsec from V18 (see Fig.\ \figfldcore), and shows light changes with frequency of $f_{\rm V13}={}$1.4505 d$^{-1}$. 
 Apparently, a small amount of the flux from V13 falls into the aperture centred on V18.
 Thus $f_2$ can be easily identified with a daily alias of $f_{\rm V13}$, that is 
 $f_2=f_{\rm V13}-{}$1 d$^{-1}$. After prewhitening the data with $f_1$ and $f_2$,
 we obtain residuals for which the periodogram is shown in Fig.\ \figmultivxviii{}c. 
 Other significant peaks appear at the 
 frequencies $f_3={}$25.5298 d$^{-1}$ and $f_4={}$1.9011 d$^{-1}$. The term $f_4$
 can be again explained as a signal originating from V13, because
 $f_4=2\times f_{\rm V13}-{}$1 d$^{-1}$. Frequency $f_3$ is intrinsic to the analyzed
 star, which is strongly testified by the presence of the combination term, $f_1+f_3$, 
 which can be also clearly seen in Fig.\ \figmultivxviii{}c. 
 The spectrum of the residuals obtained after
 prewhitening the observations with all detected frequencies (Fig.\ \figmultivxviii{}d)
 is practically flat with some higher signal at low frequencies that can be attributed to
 the instrumental effects. Thus, we found V18 to be a double-mode star.
 Phase diagrams of the two components, $f_1$ and $f_3$, are shown in the right panels of Fig.\
 \figmultivxviii{}. 

\begin{figure}[tb]
\hbox to\hsize{\hss\includegraphics{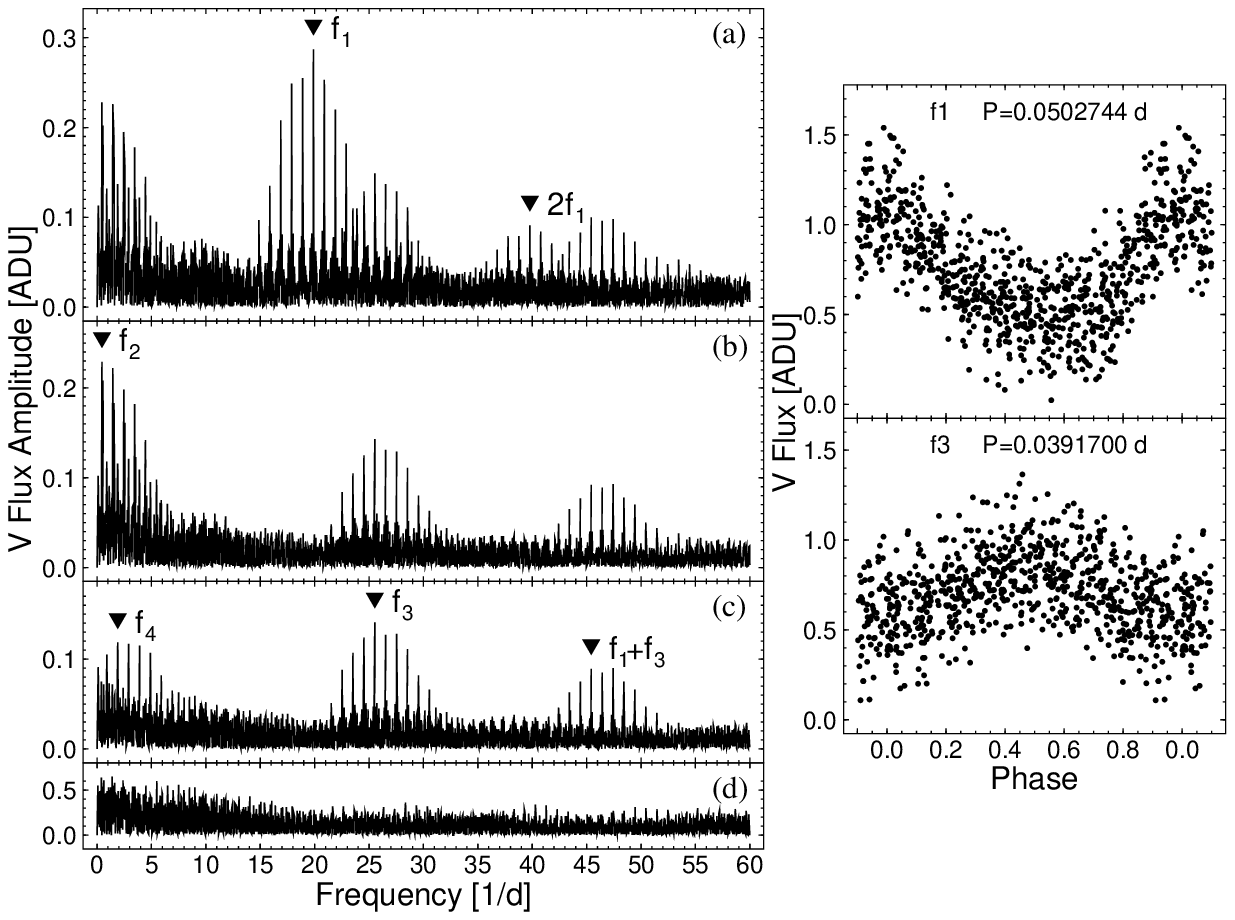}\hss}
\FigCap{%
 {\em Left panels:} Amplitude spectra of the SX Phe star V18: 
 (a) for original $V$-filter observations,
 (b) after prewhitening with frequency $f_1={}$19.8909 d$^{-1}$ and its harmonics,
 (c) after removing frequencies $f_1$ and $f_2={}$0.4478 d$^{-1}$,
 (d) after prewhitening with frequencies $f_1$, $f_2$,
 $f_3={}$25.5298 d$^{-1}$, $f_4={}$1.9011 d$^{-1}$,
 and the combination frequency $f_1+f_3$.
 The ordinate scale is the same in all panels.
 {\em Right panels:} Separated light curves of V18 for the $f_1$ and $f_3$
 components (from {\em top\/} to {\em bottom\/}) phased with appropriate periods.
 Arbitrarily chosen initial epoch is the same for both light curves.
 The ordinate scale is the same in both panels.}
\end{figure}
 
 The ratio of the periods for the double-mode SX Phe star V18 is equal to
 $P_3/P_1={}$0.779. This value (shown as squared cross in Fig.\ \figsxpheppr) 
 agrees very well with theoretical predictions for double-mode 
 radial pulsations in intermediate metallicity SX Phe stars. We conclude, that
 V18 is a variable pulsating simultaneously in the radial fundamental
 and first overtone modes. This star will be used in the next section devoted
 to the period -- luminosity relation for SX Phe stars.

\section{Period -- Luminosity Relation for SX Phe Stars}

Although intrinsically fainter than Cepheids and RR Lyr stars, SX Phe
variables are gaining increasing attention because they can be used
as distance indicators.
They are short-period pulsating
stars considered as a Population II counterparts of $\delta$ Sct stars and are found
frequently in globular clusters. The existence of the period -- luminosity relation for SX Phe stars
has been clearly established both theoretically (Santolamazza et al.\ 2001) and 
empirically (e.g.\ McNamara 1995, Poretti et al.\ 2008). 

SX Phe stars, however, are known to pulsate both in radial and/or nonradial modes.
The period -- luminosity relation should be defined for stars pulsating in the same
radial mode.
First, we need to assign oscillation modes to periodicities observed in SX Phe variables. 
Unfortunately, there is no unique method of mode identification based on a simple photometric
parameters derived from the light curve alone. 
This problem can be solved with the usage of SX Phe stars pulsating simultaneously in 
two radial modes, because their period ratios fall in a narrow range, 0.77${}-{}$0.79, 
for the first overtone (FO) and fundamental (F) radial modes, and 0.81${}-{}$0.83
for the second (SO) and first overtones (Santolamazza {\em et al.\/}\ 2001). 
This approach was recently used 
by Cohen and Sarajedini (2012) and Kopacki and Pigulski (2012b).
Unfortunately, the number of known double-mode variables of this type 
is rather small.

We performed an extensive literature search for SX Phe stars in Galactic globular clusters,
supplemented with the catalogue of SX Phe stars in globular clusters by
Rodr\'\i{}guez and L\'opez-Gonz\'alez (2000) and revised
version of the CVSGC by Clement et al.\ (2001). Altogether, 263 stars of this type 
with suitable data (pulsation period(s) and mean $V$ brightness) 
located in 30 globular clusters were found. This sample includes 27 radial double-mode pulsators.
Among them, 23 are SX Phe stars pulsating simultaneously in the fundamental and first
overtone modes (F+FO) and the other four are exhibiting oscillations in the
first and second overtones (FO+SO).

The basic parameters of radial double-mode SX Phe variables are given in Table \tabsxphedm.
For each star we provide the name of the parent cluster, star designation, 
mean $V$ magnitude, $\langle V\rangle$, primary and secondary periods, $P_1$ and $P_2$,
period ratio, $P_2/P_1$, and identification of excited modes.
We included also the double-mode SX Phe variable
discovered recently in NGC\,4833 by Kopacki (2013), although its standard $V$ magnitude is not
known at present with a sufficient accuracy (Darragh and Murphy 2012).
The observed period ratios for radial double-modes SX Phe stars, together
with the theoretical values for the first three radial
modes taken from Santolamazza {\em et al.\/} (2001),
are shown as a function of the primary (F for F+FO or FO for FO+SO stars) period in Fig.\ \figsxpheppr.
It is interesting to note, that most of the
F+FO pulsators have fundamental periods in a rather narrow range between 0.047 d and 0.066 d.

\begin{table}[htb]
 \TabCap{12.5cm}{Radial double-mode SX Phe stars in Galactic globular clusters}
 \TableFont
 \def\hst{\hbox to0pt{$^\star$\hss}} 
 \def\bn{\setbox0=\hbox{0}\hskip\wd0}
 \def\mp{\omit\hfil--\hfil}
 \hbox to\hsize{\hfill\vbox{\tabskip=2pt 
  \halign{%
   #\hfil\tabskip=9pt&
   #\hfil&
   #\hfil&
   #\hfil&
   #\hfil&
   \hfil#\hfil&
   #\hfil\tabskip=2pt\cr 
   \noalign{\hrule\vskip3pt}
   Cluster&
   Var& 
   $\langle V\rangle$& 
   $P_1$& 
   $P_2$& 
   \omit\hfil$P_2/P_1$\hfil& 
   Ident\cr
   \noalign{\vskip1pt}
   && [mag]& [d]& [d]& & \cr
   \noalign{\vskip3pt\hrule\vskip3pt}
 47 Tuc& V1& 15.59& 0.0633& 0.049375& 0.780& F+FO\cr
 47 Tuc& V2& 14.86& 0.10187& 0.07884& 0.774& F+FO\cr
 M\,79& V18& 18.255& 0.0502744& 0.0391700& 0.779& F+FO\cr
 NGC\,3201& V2& 17.37& 0.03726& 0.02911& 0.781& F+FO\cr
 NGC\,3201& V7& 16.35& 0.06730& 0.052459& 0.779& F+FO\cr
 NGC\,3201& V8& 17.05& 0.05428& 0.042339& 0.780& F+FO\cr
 NGC\,4833& V31& --& 0.0533323& 0.041726& 0.782& F+FO\cr
 M\,53& V99& 18.456& 0.0564200& 0.0441200& 0.782& F+FO\cr
 M\,53& V101& 19.211& 0.0525200& 0.0410700& 0.782& F+FO\cr
 NGC\,5053& NC13& 19.583& 0.03396& 0.02680& 0.789& F+FO\cr
 $\omega$ Cen& V194& 17.016& 0.0471777& 0.0368144& 0.780& F+FO\cr
 $\omega$ Cen& V204& 16.881& 0.0493757& 0.0382333& 0.774& F+FO\cr
 $\omega$ Cen& V220& 16.986& 0.0528868& 0.0411250& 0.778& F+FO\cr
 $\omega$ Cen& V225& 16.845& 0.0486381& 0.0378580& 0.778& F+FO\cr
 $\omega$ Cen& V249& 17.435& 0.0349468& 0.0285714& 0.818& FO+SO\cr
 $\omega$ Cen& NV305& 17.384& 0.0441445& 0.0365672& 0.828& FO+SO\cr
 $\omega$ Cen& NV322& 17.096& 0.0479562& 0.0372552& 0.777& F+FO\cr
 $\omega$ Cen& NV323& 16.638& 0.0635009& 0.0493547& 0.777& F+FO\cr
 $\omega$ Cen& NV326& 17.041& 0.0569058& 0.0443705& 0.780& F+FO\cr
 NGC\,5466& V33& 18.876& 0.0499& 0.0390& 0.782& F+FO\cr
 NGC\,5466& V35& 18.986& 0.0505& 0.0395& 0.782& F+FO\cr
 NGC\,5466& V36& 18.790& 0.05519& 0.04217& 0.764& F+FO\cr
 NGC\,5466& V40& 19.133& 0.0451& 0.03655& 0.810& FO+SO\cr
 NGC\,5466& V41& 19.277& 0.0386& 0.0306& 0.793& F+FO\cr
 M\,13& V46& 17.23& 0.052175& 0.041285& 0.791& F+FO\cr
 M\,92& V34& 17.13& 0.0830138& 0.0655600& 0.790& F+FO\cr
 M\,55& V41& 16.53& 0.0452& 0.0365& 0.808& FO+SO\cr
 M\,71& D6& 16.333& 0.06109& 0.04772& 0.781& F+FO\cr
  \noalign{\vskip3pt}
  \noalign{\vskip3pt\hrule}
  }}\hfill}
\end{table}

In order to plot period -- luminosity relation for the known SX Phe type stars
we need
absolute magnitudes of these stars. They were derived from the apparent distance
moduli $(m-M)_V$ of the parent globular clusters. Values of $M_V$ were
determined using metallicity -- absolute magnitude 
calibration for the mean brightness of the cluster's horizontal branch:
$$M_V({\rm HB})=0.22\times{\rm [Fe/H]}+0.89,$$ 
which is based on RR Lyr stars and was derived by Gratton {\em et al.\/} (2003). The average
visual magnitudes of the horizontal branch stars, $m_V({\rm HB})$, and cluster metallicities, [Fe/H], 
were taken from
the Catalogue of Parameters for Milky Way Globular Clusters by Harris (1996). The results of these
calculations are shown in Fig.\ \figsxpheplr.

\begin{figure}[tb]
\hbox to\hsize{\hss\includegraphics{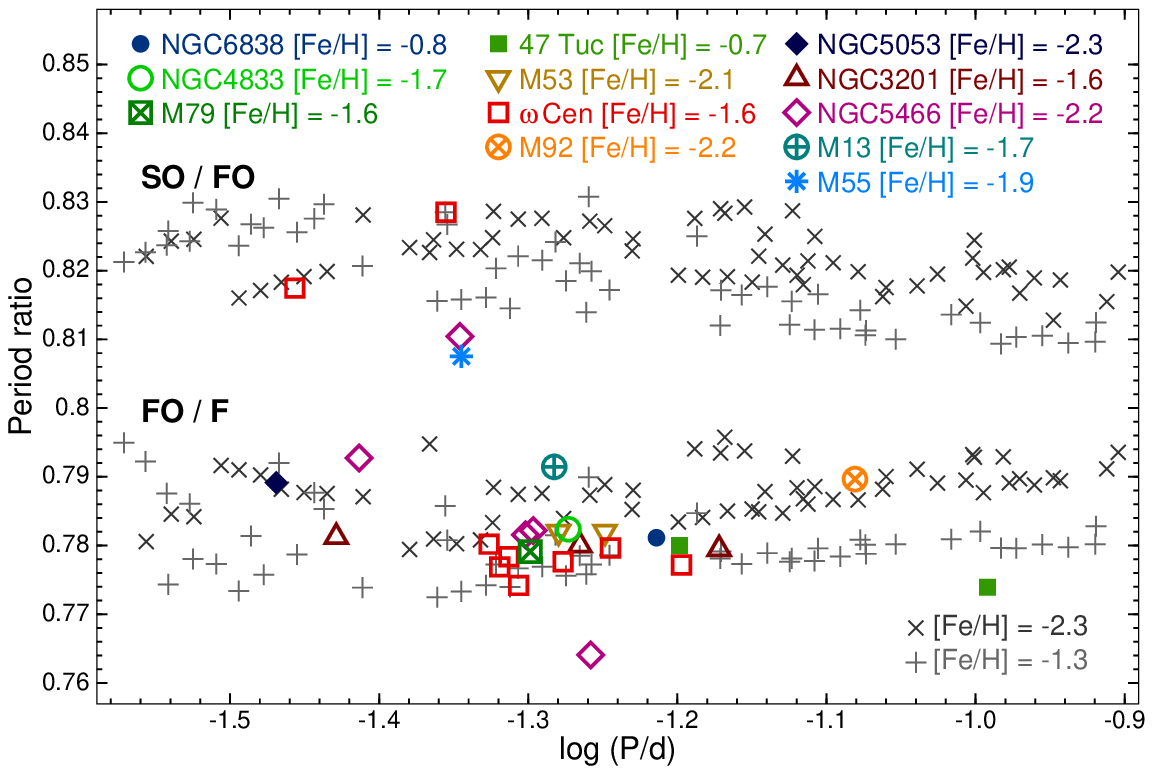}\hss}
\FigCap{%
 The generalized Petersen diagram for the fundamental (F),
 first overtone (FO) and second overtone (SO) radial modes in double-mode
 SX Phe stars. Period ratio is 
 shown as a function of period of the fundamental (or first overtone) mode. 
 Theoretical period ratios given by Santolamazza {\em et al.\/} (2001) for metallicities 
 [Fe/H]${}=-$1.3 (pluses) and [Fe/H]${}=-$2.3 (crosses) are shown for various model masses,
 luminosities and effective temperatures. We also show period ratios for all radial double-mode 
 SX Phe stars known in Galactic globular clusters, listed in Table \tabsxphedm.
}
\end{figure}

As can be seen from Fig.\ \figsxpheplr, the scatter in the period -- luminosity diagram for
SX Phe stars is very large.
The main reason for this is surely the occurrence of stars pulsating in different modes, both
radial and non-radial. It is usually assumed that SX Phe stars having ranges of variability
$\Delta V>{}$0.3 mag and showing asymmetric light curves, are pulsating in the radial fundamental
mode (e.g.\ McNamara 1995). However, the left panel of Fig.\ \figsxpheplr\ shows clearly, that these 
conditions are
rather poor discriminants of the radial pulsation mode in SX Phe stars. Dividing our sample
of SX Phe stars according to $\Delta V$ into three separate groups and comparing
them with the period -- luminosity calibrations of Cohen and Sarajedini (2012) and McNamara (1995),
we can see that even stars with ranges $\Delta V>{}$0.6 mag do not form tight relation. We can,
however, conclude that the larger light amplitude, the higher possibility 
that a given star pulsates in the fundamental radial mode.

In order to define the period -- luminosity relations for SX Phe stars, for which we know for sure they
are pulsating in radial modes, we made use of confirmed double-mode radial pulsators found
in Galactic globular clusters. The periods and absolute magnitudes of the F+FO and FO+SO double-mode 
SX Phe stars are shown in the right panel of Fig.\ \figsxpheplr. 
Now, evident separation between F and FO modes defining two parallel strips can be discerned in this figure. 
To increase the number of stars, we
included in this fit single-mode variables with high amplitudes (HA), $\Delta V>{}$0.6 mag,
obtaining totally 35 stars: 22 F+FO and 13 F pulsators. 
Fitting a straight line to the F components yields  equation:
$$ M_V=(-3.33\pm0.33)\times[\log (P_{\rm F}/{\rm d})+1.24]+(2.69\pm0.03)\hskip0.5cm \sigma=0.19{\rm\ mag},$$
%
where $\sigma$ denotes standard deviation from the fit. 
This relation is shown with a solid line in the right panel of Fig.\ \figsxpheplr.
As can be seen, the scatter around
derived relation, and consequently the error of the slope is relatively large. 
However, the uncertainity of the absolute magnitude is determined by the error of the zero point, which is 
defined above as $M_V$ at the $\log(P_{\rm F}/{\rm d})=-$1.24, and has quite a small value.

It should be noted that Cohen and Sarajedini (2012) identified fundamental radial mode SX Phe
stars as those having $\Delta V >{}$0.2 mag and $M_V$ magnitudes close to the 
fundamental-mode period -- luminosity relation derived from radial double-mode SX Phe stars only.
In view of the arguments given above, we consider this approach to be inaccurate. 
Instead, we assumed
that only SX Phe stars with $\Delta V>{}$0.6 mag are
fundamental mode pulsators. 

We also derived the following period -- luminosity relation for
FO mode using FO periods of the F+FO and FO+SO stars:
$$ M_V=(-3.39\pm0.42)\times[\log (P_{\rm FO}/{\rm d})+1.24]+(2.32\pm0.07)\hskip0.5cm \sigma=0.20{\rm\ mag},$$
%
The number of points used in the fit was 24. We excluded two the faintest stars
(FO+SO pulsators from $\omega$ Cen), which seem to be too faint for their periods 
(see the right panel of Fig.\ \figsxpheplr). The fitted relation is indicated in Fig.\ 
\figsxpheplr\ with a dashed line. 
As expected, the slopes of the two period -- luminosity relations, for fundamental
and first overtone periods, are almost the same. These values compare well with the
slope derived by Cohen and Sarajedini (2012).

\begin{figure}[tb]
\hbox to\hsize{\hss\includegraphics{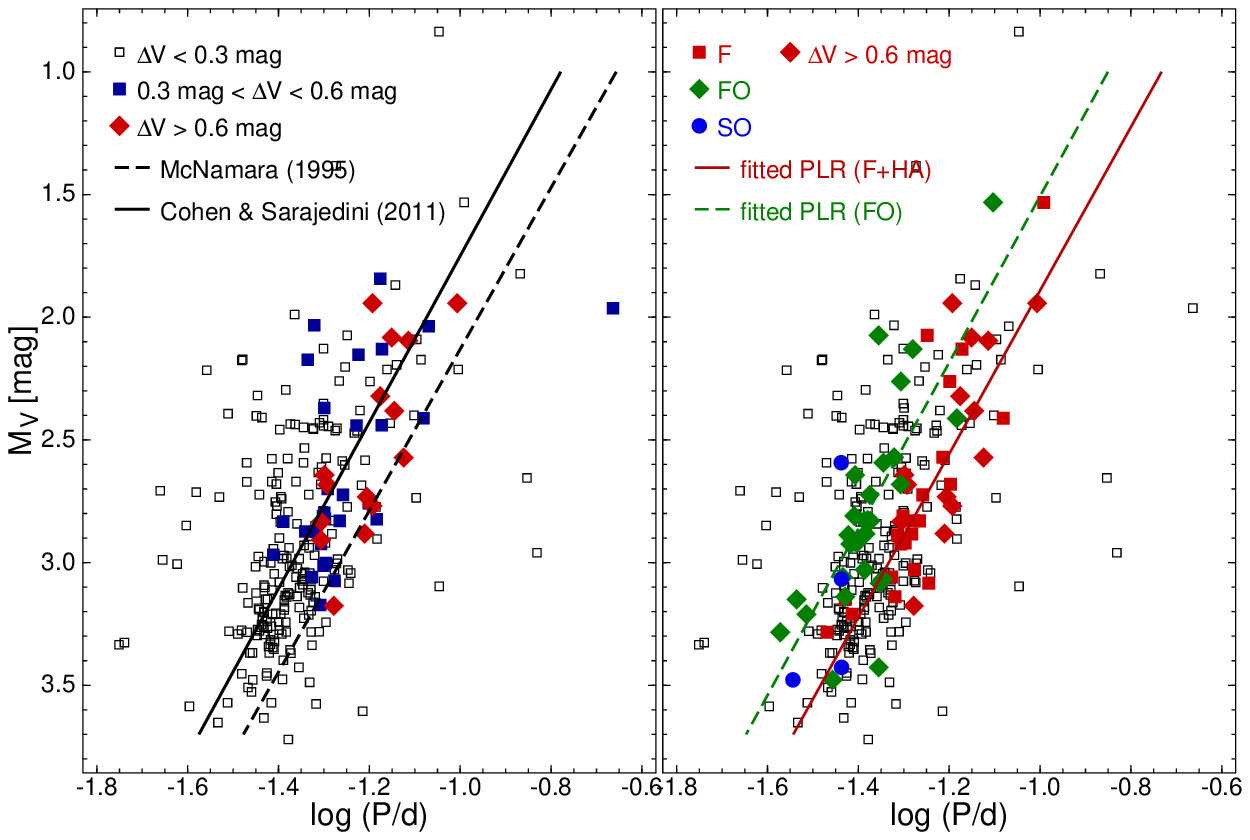}\hss}
\FigCap{%
 The period -- luminosity diagrams for analyzed SX Phe stars.
 In the {\em left panel\/} SX Phe stars with different ranges of
 variability, $\Delta V$ are shown: 
 those with $\Delta V<{}$0.3 mag with black open squares, those with
 0.3${}<\Delta V<{}$0.6 mag with blue filled squares, and those
 with $\Delta V>{}$0.6 mag with red filled diamonds. For the reference, the solid and dashed
 lines represent the period -- luminosity calibrations of 
 Cohen and Sarajedini (2012) and McNamara (1995), respectively.
 In the {\em right panel\/} fundamental (F), first and second overtone (FO and SO) periods of
 the double radial-mode SX Phe stars are indicated with red squares,
 green diamonds, and blue circles, respectively. Red diamonds represent
 high-amplitude (HA) stars with $\Delta V>{}$0.6 mag.  
 The solid red line shows period -- luminosity relation for the fundamental mode 
 obtained from the linear fit for the F+FO and high-amplitude pulsators. The dashed
 green line is a derived fit for the observed F+FO and FO+SO stars.}
\end{figure}

In order to include in a single fit as much points as possible, we also used all 
available periods of radial modes observed in SX Phe variables, namely fundamental periods of high-amplitude
and F+FO stars and fundamentalized FO and SO periods of F+FO and SO+FO pulsators.
We assumed the following mean period ratios 
$\langle P_{\rm FO}/P_{\rm F}\rangle\approx{}$0.783 and 
$\langle P_{\rm SO}/P_{\rm FO}\rangle$${}\approx{}$0.820. In total, this sample consisted
of 62 points.
From these data we determined the final
equation:
$$ M_V=(-3.35\pm0.27)\times[\log (P_{\rm F}/{\rm d})+1.24]+(2.68\pm0.03)\hskip0.5cm \sigma=0.21{\rm\ mag},$$
with coefficients very similiar to those of the period -- luminosity relation derived above
from F pulsators only.

Theoretical predictions for pulsational models of SX Phe stars show that inclusion
of color and/or metallicity terms greatly reduces the dispersion of the period --
luminosity -- color -- metallicity relation ({Petersen and Freyhammer 2002). 
We attempted to take into account the 
metallicity term in the linear least-squares fit, but found it, in agreement with Cohen and Sarajedini (2012),
insignificant.

\section{Summary and Conclusions}

 Our wide-field variability survey resulted in the discovery of 20 new variable stars
 in the field of globular cluster M\,79. Of these, one is an RR Lyr star, four (possibly five) are
 SX Phe stars, one shows W UMa-type light variations and another one is a periodic variable
 of unknown type. The remaining twelve stars are irregular red giants belonging to the
 cluster or semiregular field stars.
 Of the SX Phe stars only three are cluster members. We also found suspected variable star
 V7 to be a W Vir type star with a period of 13.985 d.
 Since we probed cluster crowded
 core with the ISM and covered large area around it, we suppose that our search revealed
 all RR Lyr stars belonging to the cluster. 

 The total number of RR Lyr stars known in M\,79 is now 11. Six of them
 are RRab stars, the remaining five are of the RRc type. With our study, we increased this number
 only by one RRc star. Nevertheless, revised mean periods of RRab and RRc stars, 
 $\langle P\rangle_{\rm ab}={}$0.71 d, $\langle P\rangle_{\rm c}={}$0.34 d, respectively, and
 relative percentage of RRc stars, $N_{\rm c}/(N_{\rm ab}+N_{\rm c})={}$45 \%, confirm
 recent results of Amigo {\em et al.\/} (2011) and Kains {\em et al.\/} (2012) 
 that M\,79 belongs to the Oosterhoff II group of globular clusters. 
 
 The mean $V$ magnitude of the horizontal branch for M\,79 determined from the intensity-weighted 
 mean magnitudes of the ten RR Lyrae stars is equal to 
 $V_{\rm HB}=\langle V\rangle_{\rm RR} ={}$16.11${}\pm{}$0.03 mag. This value 
 is very close to the $V$ brightness of the cluster's horizontal branch given in the latest version of the
 Harris' (1996) catalogue, 16.15 mag. Using the RR Lyr based
 metallicity -- luminosity relation from Gratton {\em et al.\/}\ (2003) we find 
 $M_V({\rm HB})={}$0.54 mag for M\,79. By adopting a reddening of $E(B-V)={}$0.01 mag and a value 
 of total-to-selective extinction
 $R_V={}$3.1, the true distance modulus is equal to $(m-M)_0={}$15.54 mag, corresponding
 to a distance of 12.8 kpc. These values fall in the range of modulus and distance estimates
 published till now and summarized by Kains {\em et al.\/} (2012).
 
 Among RR Lyr stars we found two variables, V4 and V9, exhibiting Blazhko effect. The RRc star V9
 shows in the frequency domain three closely spaced components, one of which corresponds 
 to the radial overtone. The other two are thus probably nonradial modes.
 One SX Phe star, V18, is a double-mode pulsator with the period ratio indicating
 pulsations in two radial modes, fundamental and first overtone.
 
 Prompted by the discovery of radial double-mode SX Phe star in M\,79, we studied
 period -- luminosity relation for this type of pulsating variables. Using 62 
 fundamental and fundamentalized periods of radial double-mode and high-amplitude SX Phe stars known in 
 Galactic globular clusters, we derived a slope and zero point of this
 relation to be, $-$3.35${}\pm{}$0.27 and 2.68${}\pm{}$0.03 mag (at $\log(P/{\rm d})=-$1.24), respectively. 
 Further searches for radial double-mode SX Phe variables are needed for a
 better determination of period -- luminosity calibration for SX Phe stars.
 
 It should be noted that the period -- luminosity relation we derived for SX Phe stars
 is based on the calibration of the absolute magnitude of cluster's horizontal
 branches and thus on RR Lyr stars. We anticipate that the Gaia mission
 (Lindegren and Perryman 1996) will
 provide more accurate distances for many field SX Phe stars, which surely allows 
 independent determination of this relation.

\Acknow{%
We acknowledge the support from the NCN grant no.\ 2011/03/B/ST9/02667.
This paper uses observations made at the South African Astronomical Observatory,
Republic of South Africa, and at 
the Siding Spring Observatory, Australia.
Some of the data presented in this paper were also obtained from the 
Mikulski Archive for Space Telescopes
(MAST). STScI is operated by the Association of Universities for 
Research in Astronomy, Inc., under NASA contract NAS5-26555. 
Support for MAST for non-HST data is provided by the NASA 
Office of Space Science via grant NAG5-7584 and by other 
grants and contracts.
}


\begin{references}
\refitem{Alard, C.}{2000}{A\&AS}{144}{363}
\refitem{Alard, C., and Lupton, R.H.}{1998}{ApJ}{503}{325}
\refitem{Amigo, P., Catelan, M., Stetson, P.B., Smith, H.A., Cacciari, C., Zoccali, M.}{2011}%
{RR Lyrae Stars, Metal-Poor Stars, and the Galaxy, Astrophysics Series}{5}{127}
\refitem{Bailey, S.I.}{1902}{Har.\ Ann.}{38}{1}
\refitem{Benk\H{o}, J.M., {\em et al.}}{2010}{MNRAS}{409}{1585}
\refitem{Clement, C.M., and Rowe, J.}{2000}{AJ}{120}{2579}
\refitem{Clement, C.M., {\em et al.}}{2001}{AJ}{122}{2587}
\refitem{Cohen, R.E., and Sarajedini, A.}{2012}{MNRAS}{419}{342}
\refitem{Dalessandro, E., Salaris, M., Ferraro, F.R., Mucciarelli, A., and Cassisi, S.}{2013}{MNRAS}{430}{459}
\refitem{Darragh, A.N., and Murphy, B.W.}{2012}{JSARA}{6}{72}
\refitem{Dolphin, A.E.}{2000a}{PASP}{112}{1383}
\refitem{Dolphin, A.E.}{2000b}{PASP}{112}{1397}
\refitem{Gratton, R.G., Carretta, E., Bragaglia, A.}{2012}{Astron.\ Astrophys.\ Rev.}{20}{50}	
\refitem{Gratton, R.G., Bragaglia, A., Carretta, E., Clementini, G., Desidera, S., Grundahl, F., and Lucatello, S.}{2003}{A\&A}{408}{529}
\refitem{Harris, W.E.}{1996}{AJ}{112}{1487}
\refitem{Jerzykiewicz, M., Pigulski, A., Kopacki, G., Mialkowska, A., and Niczyporuk, S.}{1996}{Acta Astron.}{46}{253}
\refitem{Kains, N., Bramich, D.M., Figuera Jaimes, R., Arellano Ferro, A., Giridhar, S., and Kuppuswamy, K.}{2012}{A\&A}{548}{92}
\refitem{Kolenberg, K., {\em et al.}}{2010}{ApJ}{713}{198}
\refitem{Kopacki, G.}{2000}{A\&A}{358}{547}
\refitem{Kopacki, G.}{2001}{A\&A}{369}{862}
\refitem{Kopacki, G.}{2005}{Acta Astron.}{55}{85}
\refitem{Kopacki, G.}{2007}{Acta Astron.}{57}{49}
\refitem{Kopacki, G.}{2009}{AIPC}{1170}{194}
\refitem{Kopacki, G.}{2013}{Acta Astron.}{63}{91}
\refitem{Kopacki, G.}{2014}{Precision Asteroseismology, Proceedings of the IAU}{301}{441}
\refitem{Kopacki, G. Drobek, D., Ko\l{}aczkowski, Z., Po\l{}ubek, G.}{2008}{Acta Astron.}{58}{373}
\refitem{Kopacki, G., Ko\l{}aczkowski, Z., and Pigulski, A.}{2003}{A\&A}{398}{541}
\refitem{Kopacki, G., and Pigulski, A.}{2012a}{arXiv}{1211}{5463}
\refitem{Kopacki, G., and Pigulski, A.}{2012b}{arXiv}{1211}{5465}
\refitem{Kov\'acs, G.}{2009}{AIPC}{1170}{261}
\refitem{Kravtsov, V., Ipatov, A., Samus, N., Smirnov, O., Alcaino, G., Liller, W., and Alvarado, F.}%
{1997}{A\&AS}{125}{1}	
\refitem{Landolt, A.U.}{1992}{AJ}{104}{340}
\refitem{Lanzoni, B., Sanna, N., Ferraro, F.R., Valenti, E., Beccari, G., Schiavon, R.P., Rood, R.T., Mapelli, M., Sigurdsson, S.}%
{2007}{ApJ}{663}{1040}
\refitem{Lindegren, L., and Perryman, M.A.C.}{1996}{A\&AS}{116}{579}
\refitem{Martin, N., Ibata, R., Bellazzini, M., Irwin, M., Lewis, G., and Dehnen, W.}{2004}{MNRAS}{348}{12}
\refitem{Mateu, C., Vivas, A.K., Zinn, R., Miller, L.R., and Abad, C.}{2009}{AJ}{137}{4412}
\refitem{McNamara, D.H.}{1995}{AJ}{109}{1751}
\refitem{Nemec, J.M.,  Linnell Nemec, A.F., and Lutz, T.E.}{1994}{AJ}{108}{222}
\refitem{Petersen, J.O., and Freyhammer, L.M.}{2002}{ASPC}{259}{136}
\refitem{Poretti, E., et al.}{2008}{ApJ}{685}{947}
\refitem{Rodr\'\i{}guez, and L\'opez-Gonz\'alez}{2000}{AAp}{359}{597}
\refitem{Rosino, L.}{1952}{Bologna Pub.}{5}{20} 
\refitem{Santolamazza, P., Marconi, M., Bono, G., Caputo, F., Cassisi, S., and Gilliland, R.L.}{2001}{ApJ}{554}{1124}
\refitem{Stetson, P.B.}{1987}{PASP}{99}{191}
\refitem{Stetson, P.B.}{2000}{PASP}{112}{925}
\refitem{Welty, D.E.}{1985}{AJ}{90}{2555}
\refitem{Zacharias, N., Monet, D.G., Levine, S.E., Urban, S.E., Gaume, R., and Wycoff, G.L.}{2004}{Bulletin of the American Astronomical Society}{36}{1418}
\end{references}
\end{document}